
\def \stone{{\it B Decays}, edited by S. Stone (World Scientific, Singapore,
1994)}
\def\barp{{\raise.35ex\hbox{${\sss (}$}}---{\raise.35ex\hbox{${\sss )}$}}}
\def\bdbarp{\hbox{$B_d$\kern-1.4em\raise1.4ex\hbox{\barp}}}
\def\bsbarp{\hbox{$B_s$\kern-1.4em\raise1.4ex\hbox{\barp}}}
\def\ks{K_{\sss S}}
\def\kl{K_{\sss L}}
\def\rht{{\sss R}}
\def\lft{{\sss L}}
\def\dcp{D_{\sss CP}}
\def\bqbqbar{$B_q^0$-${\overline{B_q^0}}$}
\def\bdbdbar{$B_d^0$-${\overline{B_d^0}}$}
\def\bsbsbar{$B_s^0$-${\overline{B_s^0}}$}
\def\mw{M_{\sss W}}
\def\mz{M_{\sss Z}}
\def\rht{{\sss R}}
\def\lft{{\sss L}}
\def\mhplus{M_{\sss H^+}}
\def\wino{{\widetilde W}}
\def\higgsino{{\widetilde H}}
\def\zino{{\tilde Z}}
\def\photino{{\tilde\gamma}}



\font\titlefont = cmr10 scaled\magstep 4
 2
\font\sectionfont = cmr10
\font\littlefont = cmr5
\font\eightrm = cmr8 

\def\ss{\scriptstyle} 
\def\sss{\scriptscriptstyle} 

\newcount\tcflag
\tcflag = 0  

\ifnum\tcflag = 0 \magnification = 1200 \fi  

\global\baselineskip = 1.2\baselineskip
\global\parskip = 4pt plus 0.3pt
\global\abovedisplayskip = 18pt plus3pt minus9pt
\global\belowdisplayskip = 18pt plus3pt minus9pt
\global\abovedisplayshortskip = 6pt plus3pt
\global\belowdisplayshortskip = 6pt plus3pt


\def\endignore{}
\def\ignore #1\endignore{}

\newcount\dflag
\dflag = 0


\def\monthname{\ifcase\month
\or January \or February \or March \or April \or May \or June%
\or July \or August \or September \or October \or November %
\or December
\fi}

\newcount\dummy
\newcount\minute  
\newcount\hour
\newcount\localtime
\newcount\localday
\localtime = \time
\localday = \day

\def\advanceclock#1#2{ 
\dummy = #1
\multiply\dummy by 60
\advance\dummy by #2
\advance\localtime by \dummy
\ifnum\localtime > 1440 
\advance\localtime by -1440
\advance\localday by 1
\fi}

\def\settime{{\dummy = \localtime%
\divide\dummy by 60%
\hour = \dummy
\minute = \localtime%
\multiply\dummy by 60%
\advance\minute by -\dummy
\ifnum\minute < 10 
\xdef\spacer{0} 
\else \xdef\spacer{} 
\fi %
\ifnum\hour < 12 
\xdef\ampm{a.m.} 
\else 
\xdef\ampm{p.m.} 
\advance\hour by -12 %
\fi %
\ifnum\hour = 0 \hour = 12 \fi
\xdef\timestring{\number\hour : \spacer \number\minute%
\thinspace \ampm}}}



\def\endtitle{}
\def\title#1\endtitle{\vskip.5in\titlefont
\global\baselineskip = 2\baselineskip
#1\vskip.4in
\baselineskip = 0.5\baselineskip\rm}
 
\def\endauthors{}
\def\authors#1\endauthors{#1}

\def\endabstract{}
\def\abstract#1\endabstract{\vskip .3in%
\centerline{\sectionfont\bf Abstract}%
\vskip .1in
\noindent#1}

\def\nopageonenumber{\footline={\ifnum\pageno<2\hfil\else
\hss\tenrm\folio\hss\fi}}  

\newcount\nsection
\newcount\nsubsection

\def\section#1{\global\advance\nsection by 1
\nsubsection=0
\bigskip\noindent\centerline{\sectionfont \bf \number\nsection.\ #1}
\bigskip\rm\nobreak}

\def\subsection#1{\global\advance\nsubsection by 1
\bigskip\noindent\sectionfont \sl \number\nsection.\number\nsubsection)\
#1\bigskip\rm\nobreak}

\def\topic#1{{\medskip\noindent $\bullet$ \it #1:}} 
\def\endtopic{\medskip}

\def\appendix#1#2{\bigskip\noindent%
\centerline{\sectionfont \bf Appendix #1.\ #2}
\bigskip\rm\nobreak}


\newcount\nref
\global\nref = 1

\def\therefs{} 


\def\ref#1#2{\xdef #1{[\number\nref]}
\ifnum\nref = 1\global\xdef\therefs{\item{[\number\nref]} #2\ }
\else
\global\xdef\oldrefs{\therefs}
\global\xdef\therefs{\oldrefs\vskip.1in\item{[\number\nref]} #2\ }%
\fi%
\global\advance\nref by 1
}

\def\listrefs{\vfill\eject\section{References}\therefs}


\newcount\nfoot
\global\nfoot = 1

\def\foot#1#2{\xdef #1{(\number\nfoot)}
\footnote{${}^{\number\nfoot}$}{\eightrm #2}
\global\advance\nfoot by 1
}


\newcount\nfig
\global\nfig = 1
\def\thefigs{} 

\def\figure#1#2{\xdef #1{(\number\nfig)}
\ifnum\nfig = 1\global\xdef\thefigs{\item{(\number\nfig)} #2\ }
\else
\global\xdef\oldfigs{\thefigs}
\global\xdef\thefigs{\oldfigs\vskip.1in\item{(\number\nfig)} #2\ }%
\fi%
\global\advance\nfig by 1 } 

\def\fig#1{\xdef #1{(\number\nfig)}
\global\advance\nfig by 1 } 


\newcount\ntab
\global\ntab = 1

\def\table#1{\xdef #1{\number\ntab}
\global\advance\ntab by 1 } 


\newcount\cflag
\newcount\nequation
\global\nequation = 1
\def\eqlabel{(1)}

\def\nexteqno{\ifnum\cflag = 0
\global\advance\nequation by 1
\fi
\global\cflag = 0
\xdef\eqlabel{(\number\nequation)}}

\def\lasteqno{\global\advance\nequation by -1
\xdef\eqlabel{(\number\nequation)}}

\def\label#1{\xdef #1{(\number\nequation)}
\ifnum\dflag = 1
{\escapechar = -1
\xdef\draftname{\littlefont\string#1}}
\fi}

\def\clabel#1#2{\xdef\eqlabel{(\number\nequation #2)}
\global\cflag = 1
\xdef #1{\eqlabel}
\ifnum\dflag = 1
{\escapechar = -1
\xdef\draftname{\string#1}}
\fi}

\def\cclabel#1#2{\xdef\eqlabel{#2)}
\global\cflag = 1
\xdef #1{\eqlabel}
\ifnum\dflag = 1
{\escapechar = -1
\xdef\draftname{\string#1}}
\fi}


\def\eeq{}

\def\eqnn #1\eeq{$$ #1 $$}

\def\eq #1\eeq{
\ifnum\dflag = 0
{\xdef\draftname{\ }}
\fi 
$$ #1
\eqno{\eqlabel \rlap{\ \draftname}} $$
\nexteqno}







\def\eqa #1\eeq{
\ifnum\dflag = 0
{\xdef\draftname{\ }}
\fi 
$$ \eqalignno{ #1 } $$
\global\cflag = 0}


\def\ie{{\it i.e.\/}}
\def\eg{{\it e.g.\/}}

\def\etal{{\it et.al.\/}}


\def\ijmp#1#2#3{{\it Int.\ J.\ Mod.\ Phys.} {\bf A#1} (19#2) #3}

\def\mpla#1#2#3{{\it Mod.\ Phys.\ Lett.} {\bf A#1}, (19#2) #3}

\def\npb#1#2#3{{\it Nucl.\ Phys.} {\bf B#1} (19#2) #3}
\def\plb#1#2#3{{\it Phys.\ Lett.} {\bf #1B} (19#2) #3}

\def\prd#1#2#3{{\it Phys.\ Rev.} {\bf D#1} (19#2) #3}

\def\prep#1#2#3{{\it Phys.\ Rep.} {\bf C#1} (19#2) #3}
\def\prl#1#2#3{{\it Phys.\ Rev.\ Lett.} {\bf #1} (19#2) #3}

\def\zpc#1#2#3{{\it Zeit.\ Phys.} {\bf C#1} (19#2) #3}


\global\nulldelimiterspace = 0pt



\def\frac#1#2{{{#1} \over {#2}}\,}  



\def\Dsl{\hbox{/\kern-.6700em\it D}} 
\def\dsl{\hbox{/\kern-.5300em$\partial$}}
\def\pxpsl{\hbox{/\kern-.5600em$p$}}
\def\ssl{\hbox{/\kern-.5300em$s$}}
\def\epssl{\hbox{/\kern-.5100em$\epsilon$}}
\def\delsl{\hbox{/\kern-.6300em$\nabla$}}
\def\lxpsl{\hbox{/\kern-.4300em$l$}}
\def\elxpsl{\hbox{/\kern-.4500em$\ell$}}
\def\kxpsl{\hbox{/\kern-.5100em$k$}}
\def\qxpsl{\hbox{/\kern-.5000em$q$}}
\def\sla#1{\raise.15ex\hbox{$/$}\kern-.57em #1}



\def\roughly#1{\mathrel{\raise.3ex\hbox{$#1$\kern-.75em\lower1ex\hbox{$\sim$}}}}
\def\lsim{\roughly<}
\def\gsim{\roughly>}






\def\ssl{{\sss L}}







\def\mpla#1#2#3{{\it Mod.\ Phys.\ Lett.} {\bf A#1}, #3 (19#2)}
\def\npb#1#2#3{{\it Nucl.\ Phys.} {\bf B#1}, #3 (19#2)}
\def\plb#1#2#3{{\it Phys.\ Lett.} {\bf B#1}, #3 (19#2)}
\def\prd#1#2#3{{\it Phys.\ Rev.} {\bf D#1}, #3 (19#2)}
\def\prep#1#2#3{{\it Phys.\ Rep.} {\bf C#1}, #3 (19#2)}
\def\prl#1#2#3{{\it Phys.\ Rev.\ Lett.} {\bf #1}, #3 (19#2)}
\def\zpc#1#2#3{{\it Zeit.\ Phys.} {\bf C#1}, #3 (19#2)}

 2

\overfullrule=0pt

\newread\epsffilein 
\newif\ifepsffileok 
\newif\ifepsfbbfound 
\newif\ifepsfverbose 
\newdimen\epsfxsize 
\newdimen\epsfysize 
\newdimen\epsftsize 
\newdimen\epsfrsize 
\newdimen\epsftmp 
\newdimen\pspoints 
\pspoints=1bp 
\epsfxsize=0pt 
\epsfysize=0pt 
\def\epsfbox#1{\global\def\epsfllx{72}\global\def\epsflly{72}%
 \global\def\epsfurx{540}\global\def\epsfury{720}%
 \def\lbracket{[}\def\testit{#1}\ifx\testit\lbracket
 \let\next=\epsfgetlitbb\else\let\next=\epsfnormal\fi\next{#1}}%
\def\epsfgetlitbb#1#2 #3 #4 #5]#6{\epsfgrab #2 #3 #4 #5 .\\%
 \epsfsetgraph{#6}}%
\def\epsfnormal#1{\epsfgetbb{#1}\epsfsetgraph{#1}}%
\def\epsfgetbb#1{%
%
%
\openin\epsffilein=#1
\ifeof\epsffilein\errmessage{I couldn't open #1, will ignore it}\else
%
%
 {\epsffileoktrue \chardef\other=12
 \def\do##1{\catcode`##1=\other}\dospecials \catcode`\ =10
 \loop
 \read\epsffilein to \epsffileline
 \ifeof\epsffilein\epsffileokfalse\else
%
%
 \expandafter\epsfaux\epsffileline:. \\%
 \fi
 \ifepsffileok\repeat
 \ifepsfbbfound\else
 \ifepsfverbose\message{No bounding box comment in #1; using defaults}\fi\fi
 }\closein\epsffilein\fi}%
%
%
\def\epsfclipstring{}
\def\epsfsetgraph#1{%
 \epsfrsize=\epsfury\pspoints
 \advance\epsfrsize by-\epsflly\pspoints
 \epsftsize=\epsfurx\pspoints
 \advance\epsftsize by-\epsfllx\pspoints
%
%
 \epsfxsize\epsfsize\epsftsize\epsfrsize
 \ifnum\epsfxsize=0 \ifnum\epsfysize=0
 \epsfxsize=\epsftsize \epsfysize=\epsfrsize
 \epsfrsize=0pt
%
%
 \else\epsftmp=\epsftsize \divide\epsftmp\epsfrsize
 \epsfxsize=\epsfysize \multiply\epsfxsize\epsftmp
 \multiply\epsftmp\epsfrsize \advance\epsftsize-\epsftmp
 \epsftmp=\epsfysize
 \loop \advance\epsftsize\epsftsize \divide\epsftmp 2
 \ifnum\epsftmp>0
 \ifnum\epsftsize<\epsfrsize\else
 \advance\epsftsize-\epsfrsize \advance\epsfxsize\epsftmp \fi
 \repeat
 \epsfrsize=0pt
 \fi
 \else \ifnum\epsfysize=0
 \epsftmp=\epsfrsize \divide\epsftmp\epsftsize
 \epsfysize=\epsfxsize \multiply\epsfysize\epsftmp
 \multiply\epsftmp\epsftsize \advance\epsfrsize-\epsftmp
 \epsftmp=\epsfxsize
 \loop \advance\epsfrsize\epsfrsize \divide\epsftmp 2
 \ifnum\epsftmp>0
 \ifnum\epsfrsize<\epsftsize\else
 \advance\epsfrsize-\epsftsize \advance\epsfysize\epsftmp \fi
 \repeat
 \epsfrsize=0pt
 \else
 \epsfrsize=\epsfysize
 \fi
 \fi
%
%
 \ifepsfverbose\message{#1: width=\the\epsfxsize, height=\the\epsfysize}\fi
 \epsftmp=10\epsfxsize \divide\epsftmp\pspoints
 \vbox to\epsfysize{\vfil\hbox to\epsfxsize{%
 \ifnum\epsfrsize=0\relax
 \includegraphics{#1}%
 \else
 \epsfrsize=10\epsfysize \divide\epsfrsize\pspoints
 \includegraphics{#1}%
 \fi
 \hfil}}%
\global\epsfxsize=0pt\global\epsfysize=0pt}%
%
%
 {\catcode`\%=12 \global\let\epsfpercent=
%
%
\long\def\epsfaux#1#2:#3\\{\ifx#1\epsfpercent
 \def\testit{#2}\ifx\testit\epsfbblit
 \epsfgrab #3 . . . \\%
 \epsffileokfalse
 \global\epsfbbfoundtrue
 \fi\else\ifx#1\par\else\epsffileokfalse\fi\fi}%
%
%
\def\epsfempty{}%
\def\epsfgrab #1 #2 #3 #4 #5\\{%
\global\def\epsfllx{#1}\ifx\epsfllx\epsfempty
 \epsfgrab #2 #3 #4 #5 .\\\else
 \global\def\epsflly{#2}%
 \global\def\epsfurx{#3}\global\def\epsfury{#4}\fi}%
%
%
\def\epsfsize#1#2{\epsfxsize}
%
%



\line{hep-ph/9608430 \hfill TECHNION-PH-96-37}
\rightline{UdeM-GPP-TH-96-39}
\rightline{August 1996}

\vskip .01in

\title
\centerline{New Physics in CP Asymmetries}
\centerline{and Rare B Decays}
\endtitle

\vskip .05in

\authors
\centerline{Michael Gronau${}^a$ and David London${}^{b,c}$}
\vskip .1in
\centerline{\it ${}^a$ Department of Physics}
\centerline{\it Technion -- Israel Institute of Technology, Haifa 32000,
Israel}
\vskip .1in
\centerline{\it ${}^b$ Physics Department, McGill University}
\centerline{\it 3600 University St., Montr\'eal, Qu\'ebec, Canada, H3A
2T8}
\vskip .1in
\centerline{\it ${}^c$ Laboratoire de Physique Nucl\'eaire, Universit\'e
de Montr\'eal}
\centerline{\it C.P. 6128, succ.~centre-ville, Montr\'eal, Qu\'ebec,
Canada, H3C 3J7}
\vskip .1in
\endauthors

\vskip .15in

\abstract
We review and update the effects of physics beyond the standard model on CP
asymmetries in $B$ decays. These asymmetries can be significantly altered
if there are important new-physics contributions to \bqbqbar\ mixing. This
same new physics will therefore also contribute to rare, flavor-changing
$B$ decays. Through a study of such decays, we show that it is possible to
partially distinguish the different models of new physics. 
\endabstract


\vfill\eject


\section{Introduction}

\ref\wolfenstein{L. Wolfenstein, \prl{51}{83}{1945}.}

\ref\alilon{A. Ali and D. London, DESY 96-140, UdeM-GPP-TH-96-38,
hep-ph/9607392, in Proc.\ of QCD Euroconference 96, Montpellier, France,
July 1996.}

Within the standard model (SM), CP violation is due to nonzero complex
phases in the Cabibbo-Kobayashi-Maskawa (CKM) quark mixing matrix. In the
Wolfenstein parametrization \wolfenstein, only the elements $V_{ub}$ and
$V_{td}$ have non-negligible phases:
\eq
V_{\sss CKM} = \left(
\matrix{ 1-{1\over 2}\lambda^2 & \lambda & A \lambda^3 (\rho - i \eta) \cr
- \lambda & 1-{1\over 2}\lambda^2 & A \lambda^2 \cr
A \lambda^3 (1 - \rho - i\eta) & - A \lambda^2 & 1 \cr} \right),
\eeq
where $\lambda = 0.22$ is the Cabibbo angle. The phase information of the
CKM matrix can be displayed elegantly using the so-called unitarity
triangle (Fig.~1), which follows from the orthogonality of the first and
third columns. CP violation is indicated by a nonzero area of the unitarity
triangle; to date, the only evidence for CP violation comes from
$\vert\epsilon_{\sss K}\vert$ in the kaon system.

\midinsert
\vskip -1.0truein
\centerline{\epsfxsize 3.5 truein \epsfbox {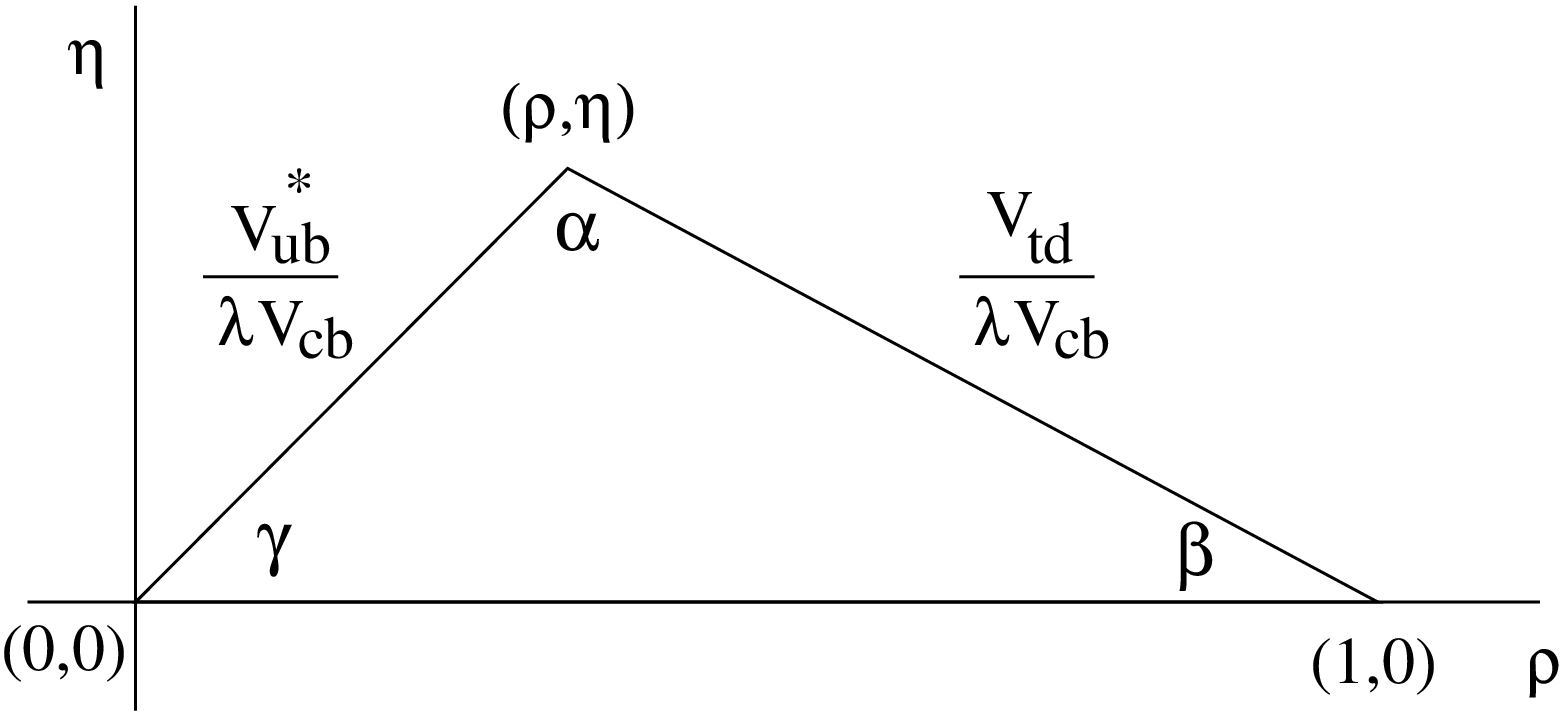}}
\vskip -1.2truein
{\eightrm Fig.~1: The unitarity triangle. The angles $\ss \alpha$, $\ss
\beta$ and $\ss \gamma$ can be measured via CP violation in the $\ss B$
system.} 
\endinsert


At present, constraints on the unitarity triangle come from a variety of
sources \alilon. The sides of the triangle can be probed directly --
$|V_{ub}/V_{cb}|$ in charmless $B$ decays, and $|V_{td}/V_{cb}|$ through
\bdbdbar\ mixing. The three angles, $\alpha$, $\beta$ and $\gamma$, are
constrained by the above measurements, as well as those of
$\vert\epsilon_{\sss K}\vert$ and $B$ decays to charmed mesons
($|V_{cb}|$). However, with the exception of $|V_{cb}|$, in all cases there
are large theoretical hadronic uncertainties in the extraction of the CKM
matrix parameters from such measurements. As such, our current knowledge of
the unitarity triangle is rather poor.

\ref\CPreview{For reviews, see, for example, Y. Nir and H. R. Quinn in
\stone, p.\ 362; I. Dunietz, {\it ibid.}, p.\ 393; M. Gronau, {\it
Proceedings of Neutrino 94, XVI International Conference on Neutrino
Physics and Astrophysics}, Eilat, Israel, May 29 -- June 3, 1994, eds. A.
Dar, G. Eilam and M. Gronau, {\it Nucl.\ Phys.\ (Proc.\ Suppl.)} {\bf B38},
136 (1995).}
In the coming years, we will be able to precisely determine the unitarity
triangle, and hence test the SM explanation of CP violation. The key
measurements involve CP-violating asymmetries in $B$ decays. Through such
measurements, the weak phases $\alpha$, $\beta$ and $\gamma$ can be
extracted with no hadronic uncertainty \CPreview, and then compared with
the SM predictions.

\ref\penguins{D. London and R. Peccei, \plb{223}{89}{257}; B. Grinstein,
\plb{229}{89}{280}; M. Gronau, \prl{63}{89}{1451}, \plb{300}{93}{163}.}

\ref\isospin{M. Gronau and D. London, \prl{65}{90}{3381}.}

It is useful at this point to review how the CP angles are probed in CP
asymmetries. Most asymmetries of interest measure mixing-induced indirect
CP violation, which comes about through the interference of the two
amplitudes $B \to f$ and $B \to {\bar B} \to f$. In order to cleanly
extract the weak phases from these CP asymmetries, two conditions must be
met. First, in the neutral $B$ system, one must have $\Gamma_{12} \ll
M_{12}$. This relation holds within the SM, where $\Gamma_{12} / M_{12}
\sim 3 \pi \, m_b^2/m_t^2 \lsim 10^{-2}$. Second, the direct decay  $B \to
f$ must be dominated by a single weak amplitude. If this is not the case,
then one may have direct CP violation, which involves unknown strong
phases. In fact, in the SM most $B$ decays which are useful for CP
asymmetries have more than one weak amplitude -- in addition to the
tree-level contribution, one may also have penguin diagrams \penguins.
However, for the cases of interest, the penguin contamination is either
unimportant or can be eliminated using isospin \isospin\ and other
considerations. We refer to Ref.~\CPreview\ for a more complete discussion
of these issues.

\ref\nirquinn{Y. Nir and H.R. Quinn, \prd{42}{90}{1473}.}

Assuming that the above two conditions are met, the CP asymmetry measures
Im$\lambda$, where $\lambda$ is a pure phase:
\eq
\label\lambdadef
\lambda = \left({X \over X^*}\right) \left({Y \over Y^*}\right)
 \left({Z \over Z^*}\right).
\eeq
The three pieces are defined as follows \nirquinn. $X$ is the weak phase of
the direct $B\to f$ decay amplitude. For example, $X_{{\bar b} \to {\bar u}
u {\bar d}} = V_{ub} V_{ud}^*$ in the SM. $Y$ is the phase of
$B^0$-${\overline{B^0}}$ mixing: \eg\ for \bdbdbar\ mixing in the SM, $Y_d
= V_{tb}^* V_{td}$. Finally, $Z$ is the phase of $K$-${\bar K}$ mixing,
which is important only if the final state $f$ contains a neutral kaon. In
the SM, assuming that $K$-${\bar K}$ mixing is dominated by box diagrams
with virtual $c$ quarks, $Z = V_{cd}V_{cs}^*$, which is real to a good
approximation in the Wolfenstein parametrization.

\ref\ADK{R. Aleksan, I. Dunietz and B. Kayser, \zpc{54}{92}{653}.}

\ref\growyler{M. Gronau and D. Wyler, \plb{265}{91}{172}. See also M.
Gronau and D. London, \plb{253}{91}{483}; I. Dunietz, \plb{270}{91}{75}.}
\ref\AKL{R. Aleksan, B. Kayser and D. London, \prl{73}{94}{18}. The effects
of new physics on these small angles are discussed in J.P. Silva and L.
Wolfenstein, CFNUL-96-09, hep-ph/9610208.}

{}From the above equation, it is straightforward to establish which CP
angles are measured in different CP asymmetries. For example, the CP
asymmetries in $\bdbarp\to\pi^+\pi^-$ and $\bdbarp\to\Psi\ks$ probe $\sin
2\alpha$ and $\sin 2\beta$, respectively. And the angle $\gamma$ can be
extracted from the CP asymmetry in $\bsbarp\to D_s^\pm K^\mp$ \ADK\ (the
function in this case is $\sin^2 \gamma$). Another way of measuring
$\gamma$, which doesn't involve mixing-induced CP violation, is via the
asymmetry in $B^\pm \to\dcp\ K^\pm$ \growyler. In all cases, the CP phases
can be obtained with no hadronic uncertainty. Of course, there are many
other CP asymmetries which can be used to obtain the angles $\alpha$,
$\beta$ and $\gamma$. (For certain decays (\eg\ $\bsbarp \to \Psi\phi$) CP
asymmetries probe very small angles of other unitarity triangles, which are
almost flat \AKL.)

\ref\nirsarid{J.M. Soares and L. Wolfenstein, \prd{47}{93}{1021};
Y. Nir and U. Sarid, \prd{47}{93}{2818}.}
\ref\alphagamma{A.S. Dighe, M. Gronau and J.L. Rosner, \prd{54}{96}{3309}.}

Once these angles are measured, it will be possible to test the SM by
comparing the measured values with the SM predictions, as well as with the
angles expected from independent measurements of the sides of the unitarity
triangle. (As we will see below, these two comparisons are not necessarily
equivalent.) As mentioned above, the SM predictions are not very precise at
present. Even so, the experimental data do somewhat constrain the CP angles
\alilon. For example, $\sin 2\beta$ must be between 0.32 and 0.94 at
95\% c.l. In addition, the predictions for $\alpha$, $\beta$ and $\gamma$
are correlated, since there is only a single complex phase in the CKM
matrix and the three angles must add up to $180^\circ$. A special
correlation was shown to exist between small values of $\sin 2\beta$ and
large values of $\sin 2\alpha$ \nirsarid, and an almost linear correlation
was found between $\alpha$ and $\gamma$ \alphagamma.

\ref\superweak{L. Wolfenstein, {\it Comments Nucl.\ Part.\ Phys.} {\bf 21},
275 (1994).}

All this presupposes that the SM is the complete description of the weak
interactions and CP violation. However, it is widely accepted that there
must be physics beyond the SM. There are a number of ways in which new
physics can manifest itself through the measurement of CP asymmetries:

\topic{(1)} The relation $\alpha+\beta+\gamma=\pi$ is violated.

\topic{(2)} Although $\alpha+\beta+\gamma=\pi$, one finds values for the CP
phases which are outside of the SM predictions.

\topic{(3)} The CP angles measured are consistent with the SM predictions,
and add up to $180^\circ$, but are inconsistent with the measurements of
the {\it sides} of the unitarity triangle.

\endtopic
\noindent
In any of these cases, it is only natural to then ask what type of new
physics could be responsible. (A special possibility is the so-called
superweak-type model \superweak, in which the CKM phase vanishes (\ie\
$\gamma=0$), and the CP asymmetries in $\bdbarp\to\pi^+\pi^-$ and
$\bdbarp\to\Psi\ks$ are equal in magnitude. It is possible, but not
necessary, that these asymmetries vanish in such models, in which case the
unitarity triangle becomes a straight line.)

\ref\DLN{C.O. Dib, D. London and Y. Nir, \ijmp{6}{91}{1253}.}

A first step in answering this question was taken in Ref.~\DLN, which
examined the effects of a variety of models of new physics on the CP
asymmetries in $B$ decays. The authors of Ref.~\DLN\ concentrated on items
{\it (1)} and {\it (2)} above. Their conclusion was that the predictions of
the SM can be considerably altered in many of these models.

\ref\grosswor{Small new-physics effects on $X$ are studied in Y. Grossman
and M.P. Worah, SLAC-PUB-7351, hep-ph/9612269.}

The reasoning goes as follows. First, it is very difficult to significantly
change the relation $\Gamma_{12} \ll M_{12}$, even in the presence of new
physics. Second, in most models of new physics, there are no new tree-level
contributions to $B$ decays.
Therefore CP asymmetries continue to measure
a well-defined CP phase, as in Eq.~\lambdadef. Of the three pieces in
Eq.~\lambdadef, only $Y$ is likely to be significantly altered by new
physics. This is because (i) although $Z$ may be modified in the presence
of new physics, only in extremely contrived models can $arg(Z)$ be changed,
and (ii) $X$ can be affected only if there are new amplitudes which can
compete with the $W$-mediated tree-level decay, and there are very few
models of new physics in which this occurs \grosswor. Thus, the principal
way that the SM predictions for CP asymmetries can be significantly
modified is if there are sizeable new-physics contributions to
$B^0$-${\overline{B^0}}$ mixing with phases different than in the SM. It is
therefore straightforward to establish, model by model, which types of new
physics can do this. Examples of models of new physics which can
significantly affect the CP asymmetries include $Z$-mediated
flavor-changing neutral currents, four generations, nonminimal
supersymmetric models, etc. In some of these models, $\vert\epsilon_{\sss
K}\vert$ is also likely to obtain sizeable new-physics contributions.

One point which was not emphasized in Ref.~\DLN\ is the third way of
detecting new physics (item {\it (3)} above). That is, even if the
new-physics contributions to $B$ mixing have the same phase as in the SM,
their presence can still be detected. This is because, although the CP
phases $\alpha$, $\beta$ and $\gamma$ are unchanged from their SM values,
the new physics affects one of the sides of the unitarity triangle, namely
the extraction of $|V_{td}/V_{cb}|$ from \bdbdbar\ mixing. Thus, the
measurements of the angles and those of the sides will be inconsistent with
one another, indicating the presence of new physics. There are several
models in which this may occur -- two-Higgs-doublet models with natural
flavor conservation, minimal supersymmetric models, etc.

\ref\nirsilv{Y. Nir and D. Silverman, \npb{345}{90}{301}.}

The question of new physics and CP asymmetries in $B$ decays has also been
discussed in Ref.~\nirsilv. This paper focussed on how new physics can
affect various relations among CP asymmetries in the SM. One of the points
made, which is of particular interest for our purposes, is the following.
Suppose that $\alpha$ (which stands for $\pi-\beta-\gamma$) and $\beta$ are
measured via CP asymmetries involving $B_d^0$ decays, and that $\gamma$ is
obtained through $B_s^0$ decays. In this case, if the phase in \bsbsbar\
mixing is identical to that of the SM, then the relation $\alpha + \beta +
\gamma = \pi$ will hold, regardless of whether there is new physics in
\bdbdbar\ or \bsbsbar\ mixing. In other words, any new-physics effects in
\bdbdbar\ mixing cancel in the sum of $\alpha$ and $\beta$.

This has important experimental implications. The angles $\alpha$ and
$\beta$ will probably be measured through CP asymmetries in
$\bdbarp\to\pi^+\pi^-$ and $\bdbarp\to\Psi\ks$, respectively. If the angle
$\gamma$ is measured through the CP asymmetry in $\bsbarp\to D_s^\pm
K^\mp$, then one might find that the three CP angles do not add up to
$180^\circ$, if there is new physics in \bsbsbar\ mixing. However, if the
angle $\gamma$ is obtained via $B^\pm \to\dcp\ K^\pm$, then, unless there
are new contributions to $B$ decays, one {\it must} find $\alpha + \beta +
\gamma = \pi$. This underlines the importance of measuring $\gamma$ (as
well as $\alpha$ and $\beta$) in a variety of independent ways. This also
demonstrates that it will be crucial to search for new-physics effects in
all three ways -- if only one of the methods {\it (1)}-{\it (3)} is used,
one might miss the presence of physics beyond the SM.

This rather lengthy introduction summarizes previous work on new physics
and CP asymmetries in the $B$ system. However, these analyses only
partially address the issue. Suppose that the CP angles $\alpha$, $\beta$
and $\gamma$ are measured, and it is found that, in fact, the presence of
new physics is indicated. Ref.~\nirsilv\ presents some tests to determine
where the new physics might be found (\eg\ \bdbdbar\ mixing, \bsbsbar\
mixing, etc.) Ref.~\DLN\ identifies which models of new physics could be
involved. However, neither of these references tells us how to distinguish
among these various models. It is this question which we address in this
paper.

As argued above, the new physics can affect the CP asymmetries mainly
through its contributions to \bdbdbar\ or \bsbsbar\ mixing, which are
flavor-changing processes. This same new physics will therefore also
affect rare flavor-changing decays, such as $b\to s X$ or $b\to d X$. (In
this paper we generically refer to such processes as ``penguin'' decays.)
This is the key point. As we will show, some models of new physics can be
distinguished by their contributions to these rare processes. In fact, for
certain models, the new-physics parameter space leading to large
contributions to $B$-${\bar B}$ mixing also predicts large deviations from
the SM predictions for certain penguin decays. Conversely, if no deviation
from the SM is found, this would so constrain the parameters of the new
physics as to render its effects in $B$-${\bar B}$ mixing, and hence the CP
asymmetries, unimportant. It is an experimental question whether or not
measurements of the rates for such penguin decays can be made before the CP
asymmetries are measured. Regardless, it is clear that measurements of CP
asymmetries and penguin decays will give complementary information. And in
fact, unless the new particles are discovered in future colliders, it will
be necessary to appeal to such measurements to infer their existence
indirectly.

The paper is organized as follows. In Section 2 we review and update the
contributions of various models of new physics to $B$-${\bar B}$ mixing,
and hence to CP asymmetries in $B$ decays. We summarize the current
experimental constraints on the new-physics parameters which determine
these contributions. For those models which can affect the SM predictions
for the CP asymmetries, in Section 3 we examine their contributions to
flavor-changing penguin decays. We conclude in Section 4.


\section{$B$-${\bar B}$ Mixing and New Physics}

\ref\Bmixing{J. Hagelin, \npb{193}{81}{123}; T. Inami and C.S. Lim, {\it
Prog.\ Theor.\ Phys.} {\bf 65}, 297 (1981), (E) {\it ibid.} {\bf 65}, 772
(1982); A.J. Buras, W. Slominski and H. Steger, \npb{238}{84}{529},
\npb{245}{84}{369}.}
\ref\DeltaMdlimit{C. Zeitnitz, invited talk at the 4th International
Workshop on $B$-Physics at Hadron Machines (BEAUTY 96), Rome, 17-21 June,
1996, and private communication.}
\ref\DeltaMslimit{C. Zeitnitz, Ref.~\DeltaMdlimit; ``Combined limit on the
$B_s^0$ oscillation frequency", contributed paper by the ALEPH
collaboration to the International Conference on High Energy Physics,
Warsaw, ICHEP96 PA08-020 (1996).}

There are a variety of models of new physics which can contribute to
\bqbqbar\ mixing ($q=d,s$), and which therefore can affect CP asymmetries
in $B$ decays. In this section we review and update the contributions of
these models to this mixing. (Note that we include several models not
discussed in Ref.~\DLN.) We also examine how the new-physics parameters are
constrained by current experimental data. In all cases, we search for new
contributions to \bqbqbar\ mixing at least comparable to that of the SM
\Bmixing:
\eq
\label\Bmix
M_{12}^{\sss SM} (B_q) =
{G_{\sss F}^2 M_{\sss B_q} \eta_{\sss B_q} \mw^2 \over 12\pi^2}
f_{\sss B_q}^2 B_{\sss B_q} x_t f_2(x_t) (V_{tq}V_{tb}^*)^2 ~,
\eeq
where the mass difference $\Delta M$ is related to $M_{12}$ by
$\Delta M_q = 2 \left\vert M_{12} (B_q) \right\vert$, $x_t=m_t^2/\mw^2$
and
\eq
\label\Bmixfun
f_2(x) = \left[ {1\over 4} + {9\over 4} \, {1\over 1-x} - {3 \over 2} \,
{1 \over (1-x)^2} - {3 \over 2} \, {x^2 \ln x \over (1-x)^3} \right]~.
\eeq

New-physics contributions to \bqbqbar\ mixing are constrained by the 
measurements of the neutral $B$-meson mass differences, $\Delta M_d$
\DeltaMdlimit\ and $\Delta M_s$ \DeltaMslimit:
\eq
\Delta M_d = (0.470 \pm 0.017) ~ps^{-1} ~,~~~~~~
\Delta M_s > 7.8 ~ps^{-1} ~.
\eeq
These values are consistent with the SM prediction [Eq.~\Bmix], and
constrain the CKM elements $V_{td}$ and $V_{ts}$ as follows \alilon:
\eq
\label\Vtdlimits
0.15 < \left\vert {V_{td} \over V_{cb}} \right\vert < 0.34~,~~~~~~
\left\vert {V_{ts} \over V_{cb}} \right\vert > 0.6 ~.
\eeq
These limits include the experimental errors on $m_t$ and $V_{cb}$, as well
as the theoretical error on $f_{\sss B_q}\sqrt{B_{\sss B_q}}$. The bounds
on these quantities due to the unitarity of the $3\times 3$ CKM matrix
alone are
\eq
\label\Vtdunitary
0.11 < \left\vert {V_{td} \over V_{cb}} \right\vert < 0.33~,~~~~~~
\left\vert {V_{ts} \over V_{cb}} \right\vert \simeq 1 ~.
\eeq

With the addition of new contributions to \bqbqbar\ mixing, the constraints
of Eq.~\Vtdlimits\ on $V_{td}$ and $V_{ts}$ are relaxed, although the
degree of relaxation is model-dependent. In certain models, the CKM matrix
remains unitary, which implies that the bounds of Eq.~\Vtdunitary\ still
hold. In other models, the $3\times 3$ CKM matrix is not unitary, so that
$V_{td}$ or $V_{ts}$ can be much smaller and in principle even vanish, in
which case \bdbdbar\ or \bsbsbar\ mixing comes entirely from new physics.

In order to see how large the new-physics contributions to \bqbqbar\ mixing
can be in specific models, it is convenient to normalize these terms by the
corresponding $W$ box-diagram terms which appear in the SM, which are 
proportional to $V^2_{tq}$. However, one should note that in some cases the
latter parameters can take values outside the SM constraints. To avoid 
confusion in this respect, we will denote the $W$-box contributions to 
\bqbqbar\ mixing by $M^{\sss W}_{12}$ rather than by $M^{\sss SM}_{12}$.

\ref\fourgen{M. Gronau and J. Schechter, \prd{31}{85}{1668};
X.G. He and S. Pakvasa, \plb{156}{85}{236};
I.I. Bigi and S. Wakaizumi, \plb{188}{87}{501};
A. Datta, E.A.Paschos and U. T\"urke, \plb{196}{87}{376};
M. Tanimoto, Y. Suetaka and K. Senba \zpc{40}{88}{539};
W.-S. Hou A. Soni and H. Steger, \plb{192}{87}441;
J.L. Hewett, \plb{193}{87}{327};
T. Hasuike, T. Hattori, T. Hayashi and S. Wakaizumi, \mpla{4}{89}{2465},
\prd{41}{90}{1691}; D. London, \plb{234}{90}{354}.}

\subsection{Four generations \fourgen}

This is a model with an additional generation of quarks and leptons,
including a new charge 2/3 quark, $t'$. The CKM matrix is $4\times 4$,
which can be parametrized by 6 angles and 3 phases. The unitarity triangle
thus now becomes a quadrangle. There are new loop-level contributions,
involving internal $t'$ quarks, to both \bqbqbar\ mixing and penguin
decays. The additional phases in the CKM matrix can play a role in the CP
asymmetries.

\ref\uppermtp{M.S. Chanowitz, M.A. Furman and I. Hinchliffe,
\plb{78}{78}{285}, \npb{153}{79}{402}.}

There is a model-independent lower bound of 45 GeV on the mass of the $t'$
coming from LEP. There are stronger constraints on $m_{t'}$ of $O(100)$ GeV
coming from hadron colliders, but these can be evaded since they depend on
how strongly the $t'$ couples to the $b$ quark. There is an upper bound of
550 GeV on $m_{t'}$ coming from partial-wave unitarity \uppermtp. A heavier
$t'$ will lead to a breakdown of perturbation theory. The strongest
constraints on the CKM matrix elements involving the $t'$ quark come from
unitarity. There are additional constraints on the $t'$ mass and its
charged-current couplings coming from the $\kl$-$\ks$ mass difference, from
$\vert\epsilon_{\sss K}\vert$, from \bdbdbar\ mixing, and from $b\to
s\gamma$. Since the measurements of all these observables agree with the
predictions of the SM, they provide upper limits on their respective $t'$
contributions, assuming no accidental cancellations. However, if one allows
for such cancellations, the constraints become correspondingly weaker.
Finally, we note that the fourth-generation neutrino must have a mass
$m_\nu > \mz/2$ due to constraints from LEP. Since this is quite unlike the
first 3 generations, many argue that the four-generation model is much less
plausible. Still, it is a logical possibility.

In this model, the extra phases in the $4\times 4$ CKM matrix enter through
the new contributions to \bqbqbar\ mixing. Assuming that this mixing is
dominated by box diagrams with $t$ and $t'$ quarks, we have
\eq
\eqalign{
M_{12}^{4-gen} (B_q) =
{G_{\sss F}^2 M_{\sss B_q} \eta_{\sss B_q} \mw^2 \over 12\pi^2}
& f_{\sss B_q}^2 B_{\sss B_q}
\left[ E(x_t,x_t)(V_{tq}V_{tb}^*)^2 \right. \cr
& \!\!\!\!\!\!\! \!\!\!\!\!\!\! \!\!\!\!\!\!\! \left. 
+~2E(x_t,x_{t^\prime})(V_{tq}V_{tb}^*)(V_{t^\prime q}V_{t^\prime b^*})
+E(x_{t^\prime},x_{t^\prime}) (V_{t^\prime q}V_{t^\prime b}^*)^2
\right]~, \cr}
\eeq
where
\eq
\eqalign{
E(x_i,x_j)=x_ix_j & \left\{\left[{1\over 4}+{3\over 2} \, {1\over (1-x_j)}
-{3\over 4} \, {1\over (1-x_j)^2}\right] {\ln x_j\over x_j-x_i}\right.\cr
& \left.~~~~~+(x_i\leftrightarrow x_j)
-{3\over 4} \, {1\over (1-x_i)(1-x_j)}\right\}~.\cr}
\eeq
Since the additional contributions to the mixing can be of a similar size
to that of the SM, but with different phases, the CP asymmetries can be
considerably altered. For example, the experimental value of \bdbdbar\
mixing can be explained in the SM if $V_{td} = 0.01$ (for $m_t=170$ GeV,
$V_{tb} = 1$); the phase of the mixing is then $arg(V_{td}V_{tb}^*)^2$. In
a 4-generation model, in which the $3\times 3$ CKM matrix is no longer
unitary, this mixing can be dominated by the fourth generation: \eg\
$V_{td} \sim 0$, $V_{t'd} = 0.005$, $V_{tb} = V_{t'b} \simeq 1/\sqrt{2}$,
and $m_{t'} = 480$ GeV. The phase of the mixing is then
$arg(V_{t'd}V_{t'b}^*)^2$, which may be quite different from the SM.

\ref\zfcnc{Y. Nir and D. Silverman, \prd{42}{90}{1477}; D. Silverman,
\prd{45}{92}{1800}; G.C. Branco, T. Morozumi, P.A. Parada and M.N. Rebelo,
\prd{48}{93}{1167}; V. Barger, M.S. Berger and R.J.N. Phillips,
\prd{52}{95}{1663}.}

\subsection{$Z$-mediated flavor-changing neutral currents \zfcnc}

In these models, one introduces an additional vector-singlet charge $-1/3$
quark, and allows it to mix with the ordinary down-type quarks. Since the
weak isospin of the exotic quark is different from that of the ordinary
quarks, flavor-changing neutral currents (FCNC's) involving the $Z$ are
induced. The $Zb{\bar d}$ and $Zb{\bar s}$ FCNC couplings, which affect $B$
decays, are parametrized by independent parameters $U_{db}$ and $U_{sb}$,
respectively, which contain new phases:
\eq
{\cal L}^{\sss Z}_{\sss FCNC} = - {g \over 2 \cos\theta_{\sss W}} \,
U_{qb} \, {\bar q}_\lft \gamma^\mu b_\lft Z_\mu ~.
\eeq

\ref\uaone{C. Albajar \etal\ (UA1 Collaboration), \plb{262}{91}{163}.}
\ref\GLN{Y. Grossman, Z. Ligeti and E. Nardi, \npb{465}{96}{369}.}

There are, however, constraints on the FCNC couplings coming from the
process $B \to \mu^+\mu^- X$. The current experimental bound on the
branching ratio of this process is \uaone\
\eq
BR(B \to \mu^+\mu^- X) < 5 \times 10^{-5}~,
\eeq
while the contributions of $Z$-mediated FCNC's to this process are
\eq
{BR(B \to \mu^+\mu^- X) \over BR(B \to \mu \nu X)} = \left[
\left( g_\lft^\mu \right)^2 + \left( g_\rht^\mu \right)^2 \right]
{ |U_{db}|^2 + |U_{sb}|^2 \over |V_{ub}|^2 + F_{ps} |V_{cb}|^2 } ~,
\eeq
where $g_\lft^\mu = -1/2 + \sin^2\theta_{\sss W}$, $g_\rht^\mu =
\sin^2\theta_{\sss W}$, and $F_{ps} \simeq 0.5$ is a phase-space factor.
The FCNC couplings $U_{qb}$, $q=d,s$ are then constrained to be
\eq
\left\vert {U_{qb} \over V_{cb}} \right\vert < 0.044 ~,
\eeq
or, taking $\left\vert V_{cb} \right\vert = 0.0388 \pm 0.0036$ \alilon,
\eq
\label\FCNClimit
\left\vert U_{qb} \right\vert < 0.0017 \pm 0.0002 ~.
\eeq
(Similar constraints can be obtained from the bound on $B\to\nu{\bar\nu} X$
\GLN.)

The $Z$-mediated flavor-changing couplings $U_{qb}$ can contribute to
\bqbqbar\ mixing:
\eq
M_{12}^{\sss Z} (B_q) =
{\sqrt{2} G_{\sss F} M_{\sss B_q} \eta_{\sss B_q} \over 12} \,
f_{\sss B_q}^2 B_{\sss B_q} (U_{qb}^*)^2 ~.
\eeq
Recall that there can be new, independent phases in $U_{qb}$.

Comparing the contribution of this new physics to \bqbqbar\ mixing
with that of the SM, we find
\eq
{\Delta M_d^{\sss Z} \over \Delta M_d^{\sss W} } =
{\sqrt{2} \pi^2 \over G_{\sss F} \mw^2} \, { 1 \over x_t f_2(x_t) }
\, { \left\vert U_{db} \right\vert^2 \over
\left\vert V_{td}V_{tb} \right\vert^2 } =
80 \, { \left\vert U_{db} \right\vert^2 \over
\left\vert V_{td} \right\vert^2 } ~,
\eeq
where we have taken $\left\vert V_{tb} \right\vert = 1$ and $m_t =
170$ GeV. 

In this model the CKM matrix is not unitary:
\eq
V^*_{ud}V_{ub} + V^*_{cd}V_{cb} + V^*_{td}V_{tb} = U_{db}~,
\eeq
so that the constraint of Eq.~\Vtdunitary\ on $V_{td}$ does not hold. Since
$|U_{db}|$ is bounded by Eq.~\FCNClimit, we find 
\eq
0.07 < \left\vert {V_{td} \over V_{cb}} \right\vert < 0.37~.
\eeq
Consequently, 
\eq
{\Delta M_d^{\sss Z} \over \Delta M_d^{\sss W} } = (0.9 {\hbox{--}} 26)
\left[ {\left\vert U_{db} / V_{cb} \right\vert \over 0.04} \right]^2 ~,
\eeq
where the numerical coefficients $0.9$ and $26$ correspond to the largest
and smallest values of $|V_{td}/V_{cb}|$. The sum of the $W$ and $Z$ 
contributions to \bdbdbar\ mixing is consistent with measurement for the
entire range of $V_{td}$. 

\bsbsbar\ mixing can be analysed similarly. However, in this case, the
effect of $U_{sb}$ on the violation of ($sb$) CKM unitarity is small, so
that
\eq
{\Delta M_s^{\sss Z} \over \Delta M_s^{\sss W} } = 0.15
\left[ {\left\vert U_{sb} / V_{cb} \right\vert \over 0.04} \right]^2 ~,
\eeq

{}From this we see that \bdbdbar\ mixing can in fact be dominated by
$Z$-mediated FCNC. And although \bsbsbar\ mixing is still mainly due to the
$W$-box contribution, the new-physics contribution may be non-negligible,
so that the new phases in $U_{sb}$ can be important. Thus, in both cases,
measurements of CP asymmetries can differ considerably from the predictions
of the SM.

\ref\seesaw{F. del Aguila and M.J. Bowick, \npb{224}{83}{107}; P.M.
Fishbane, R.E. Norton and M.J. Rivard, \prd{33}{86}{2632}; W. Buchm\"uller
and M. Gronau, \plb{220}{89}{641}.}
It is interesting to note that there exist specific models with seesaw-like
predictions for the flavor-changing $Z$ couplings \seesaw:
\eq
U_{qb} = \sqrt{{m_q m_b\over M^2}},~~~~~M={\cal O}(0.1-1~{\rm TeV})~.
\eeq
Depending on the precise value of $M$, these couplings do not lie too far
below the present limit [Eq.~\FCNClimit], and may thus give sizeable
contributions to \bdbdbar\ and \bsbsbar\ mixing.

\ref\frampton{P.H. Frampton and D. Ng, \prd{43}{91}{3034}; A.W. Ackley,
P.H. Frampton, B. Kayser and C.N. Leung, \prd{50}{94}{3560}; P.H. Frampton
and S.L. Glashow, HUTP-96/A044, hep-ph/9609496.}

In one particular flavor-changing $Z$ model \frampton, CP is violated
spontaneously, the CKM matrix is essentially real, and the new
contributions to \bqbqbar\ mixing lead to very small phases. The unitarity
triangle becomes a straight line and all CP asymmetries are expected to be
tiny.

\subsection{Multi-Higgs-doublet models}

\ref\weingla{S. Weinberg, \prl{37}{76}{657}; S.L. Glashow and S. Weinberg,
\prd{15}{77}{1958}; E.A. Paschos, \prd{15}{77}{1966}.}
\ref\leegeorgi{T.D. Lee, \prd{8}{73}{1226}, \prep{9}{74}{143}; H. Georgi,
{\it Hadronic J.} {\bf 1}, 155 (1978).} 

Models with more than one Higgs doublet can be classified into two types:
(i) models with natural flavor conservation \weingla, in which there are
no flavor-changing neutral currents, and (ii) models in which
flavor-changing interactions can be mediated by neutral scalars
\leegeorgi. We discuss these in turn.

\noindent
{\it (i) Natural Flavor Conservation}

\ref\nfhiggs{See \eg\ J.F. Gunion, H.E. Haber, G.L. Kane and S. Dawson,
{\it The Higgs Hunter's Guide} (Addison-Wesley, Reading, MA, 1990), and
references therein.}
\ref\franzini{G.G. Athanasiu, P.J. Franzini and F.J. Gilman,
\prd{32}{85}{3010}.}
%

In models with natural flavor conservation \nfhiggs, the new charged
scalars may give significant contributions to \bqbqbar\ mixing if their
masses lie in the range of 50 GeV to about 1 TeV \franzini. The Yukawa
couplings of the charged scalars to up- and down-type quarks are given by
\eq
\label\chHiggscoup
{\cal L}_{\sss H^\pm} = {g \over \sqrt{2} \mw} \sum_i \left(
X_i {\bar U}_\lft V_{\sss CKM} M_{\sss D} D_\rht +
Y_i {\bar U}_\rht M_{\sss U} V_{\sss CKM} D_\lft \right) H_i^+ + h.c.
\eeq
Here $U$ ($D$) is a vector of up-type (down-type) quarks, and $M_{\sss U}$
($M_{\sss D}$) is the diagonal charge $2/3$ (charge $-1/3$) quark mass
matrix. $X_i$ and $Y_i$ are complex coupling constants arising from the
mixing in the scalar sector. In the case of two Higgs doublets, in which
one doublet provides masses to all quarks and the other decouples from the
quark sector (model I), $Y=-X=\cot\beta$, where $\tan\beta$ is the ratio of
the two vacuum expectation values. In a more popular version of the
two-Higgs-doublet model (model II), found in supersymmetric models for
example (see discussion below), one scalar ($\phi_1$) gives mass to the
up-type quarks while the other scalar ($\phi_2$) gives mass to the
down-type quarks. In this case $X=\tan\beta$ and $Y=\cot\beta = v_2/v_1$.

\ref\grosnir{P. Krawczyk and S. Pokorski, \npb{364}{91}{10}; Y. Grossman
and Y. Nir, \plb{313}{93}{126}.}
For simplicity of presentation in the following we assume that only one
charged Higgs is light; the others are heavy and decouple. This leaves two
complex coupling constants, $X$ and $Y$. There are several useful
observations regarding Eq.~\chHiggscoup. First, the $Y$ term is dominated
by the $t$-quark: ${\cal L}_{\sss H^\pm}^{\sss Y} \sim Y \, (m_t/\mw) \,
H^+ \, {\bar t}_\rht V_{ti} d_{i\lft}$. Second, due to the smallness of the
down-type quark masses, the $X$ term is important only if $|X| \gg |Y|$. In
this region of parameter space, CP violation can appear in charged-Higgs
exchange if $X$ and $Y$ have a nonzero relative phase. However, the
observed rate of $b \to s\gamma$ constrains ${\rm Im}(XY^*) < 2$-4
\grosnir, thus ruling out the possibility that CP-violating effects due to
charged-Higgs exchange can compete with those due to $W$ exchange. Since
the inclusion of the $X$ term does not lead to new CP violation, and since
in general it is much smaller than the $Y$ term, from here on we will
generally ignore the $X$ term altogether. We will refer to the possibility
of very large $X$ only when its effect is particularly important.

\ref\abbott{L.F. Abbott, P. Sikivie and M.B. Wise, \prd{21}{80}{1393}.}
There are two types of box diagrams involving charged Higgs bosons which
contribute to \bqbqbar\ mixing: those with one $H$ and one $W$, and those
with two $H$'s. The total charged-Higgs contribution is given by \abbott
\eq
M^{{\sss H}^+}_{12}(B_q) =
{G_{\sss F}^2 M_{\sss B_q} \eta_{\sss B_q} \mw^2 \over 48\pi^2}
f_{\sss B_q}^2 B_{\sss B_q} (V_{tq}V_{tb}^*)^2
\left[ I_{\sss HH} + I_{\sss HW} \right],
\eeq
where
\eq
I_{\sss HH} = x_t \, y_t \, I_1(y_t) |Y|^4 ~~,~~~~
I_{\sss HW} = 2 \, x_t \, y_t [4I_2(x_t,y_t) + I_3(x_t,y_t)] |Y|^2 ~~,
\eeq
with
\eq
I_1(y) = {1 + y\over (1 - y)^2} + {2y\ln y\over (1 - y)^3} ~~,
\eeq
\eq
4I_2(x,y) + I_3(x,y) = {(x-4y)\ln y\over (y-x)(1-y)^2} + {3x\ln x\over
(y-x)(1-x)^2} + {x-4\over (1-x)(1-y)} ~.
\eeq
Here, $x_q \equiv m_q^2/\mw^2$ and $y_q \equiv m_q^2/\mhplus^2$. Note that
$M^{{\sss H}^+}_{12}(B_q)$ involves the same CKM factors and has the same
phase as the $W$-box contributions. Therefore, the ratio of the two terms
is independent of CKM factors and is positive, such that the two 
contributions add up constructively. For $\mhplus$ in the range 100-400
GeV, this ratio may be approximated to within about 20\% by an inversely
linear relation:
\eq
\label\chHiggsratio
{M^{{\sss H}^+}_{12}(B_q)\over M^{\sss W}_{12}(B_q)} \approx
\left({100~{\rm GeV} \over \mhplus} \right) \, \left[
0.24 \, |Y|^4 + 1.05 |Y|^2 \right]~.
\eeq
For smaller and larger Higgs masses this simplified expression holds within
about 30\%.

\ref\CLEObsgamma{M.S. Alam \etal\ (CLEO Collaboration),
\prl{74}{95}{2885}.}
\ref\Alireview{For a review, including references to the literature, see A.
Ali, DESY 96-106, hep-ph/9606324.}
\ref\grossman{Y. Grossman, \npb{426}{94}{355}.}
\ref\btosgamma{J.L. Hewett, T. Takeuchi and S. Thomas, SLAC-PUB-7088,
hep-ph/9603391, to appear in {\it Electroweak Symmetry Breaking and Beyond
the Standard Model}, edited by S. Dawson, H.E. Haber and S. Siegrist (World
Scientific, 1996), and references therein.}

The parameters $|Y|$ and $\mhplus$ are constrained by comparing the
observed rate of $b\to s\gamma$ \CLEObsgamma\ with the SM prediction
\Alireview. These constraints also depend on $X$ \grossman. At $3\sigma$
the bounds are
\eq
\label\chHiggslimit
-0.56 < |Y| {1\over 3}G_{\sss W}(y_t) + XY^* G_{\sss H}(y_t) < 0.27~,
\eeq
where
\eq
\label\GWdef
\eqalign{
G_{\sss W}(y) & = {y\over 12(1-y)^4}[(7-5y-8y^2)(1-y) + 6y(2-3y)ln(y)]~,
\cr 
G_{\sss H}(y) & = {y\over 6(y-1)^3}[(3-5y)(1-y) + 2(2-3y)ln(y)]~. \cr}
\eeq
The implications of these bounds on $Y$ and $\mhplus$ depend somewhat on
the details of the model \btosgamma. In a two-Higgs-doublet model of type
II ($XY=1$), charged-Higgs masses below about 300 GeV are already excluded,
independent of the value of $Y$. In a two-Higgs-doublet model of type I
($X=-Y$), Higgs masses in the entire range $\mhplus < 800$ GeV are 
excluded if one assumes $|Y| > 2.7$. However, in a general 
multi-Higgs-doublet model, in which $X$ and $Y$ are independent parameters,
the constraints become weaker. This leaves a large region of
$|Y|$--$\mhplus$ parameter space in which the charged-Higgs contribution to
\bqbqbar\ mixing [Eq.~\chHiggsratio] can be significant and even dominant.
For example, in a general model, the values $\mhplus=400$ GeV and $Y=3$ are
allowed, which implies $M^{{\sss H}^+}_{12}/M^{\sss W}_{12} =7$.  In this
case the sum of the $W$ and $H^+$ terms (dominated by $H^+$) is consistent
with the measurement of \bdbdbar\ mixing for the smallest values of
$V_{td}$ in the unitarity range [Eq.~\Vtdunitary].

\ref\tomalin{I. Tomalin (ALEPH Collaboration), invited talk at the
International Conference on High Energy Physics, Warsaw, ICHEP96 (1996).} 
\ref\Grant{Sizeable neutral-Higgs contributions to $R_b$ occur only for
very large values of $X$, see A.K. Grant, \prd{51}{95}{207}.}
\ref\Rbpaper{For a discussion of new contributions to $R_b$ in the context
of multi-Higgs-doublet models, see, for example, M. Boulware and D.
Finnell, \prd{44}{91}{2054}; Y. Grossman, \npb{426}{94}{355}; A.K. Grant,
Ref.~\Grant; P. Bamert, C.P. Burgess, J.M. Cline, D. London and E. Nardi,
McGill-96/04, UdeM-GPP-TH-96-34, WIS-96/09/Feb-PH, hep-ph/9602438, to
appear in Phys.\ Rev.\ {\bf D}.} 

There is also a bound on the parameters $|Y|$ and $\mhplus$ from the latest
ALEPH measurement of $R_b \equiv \Gamma(Z \to b{\bar b})/\Gamma(Z \to
{\hbox{hadrons}}) = 0.2158 \pm 0.0014$ \tomalin. Assuming the neutral-Higgs
contribution to $R_b$ is small (\ie\ $X$ is not too large) \Grant, at
3$\sigma$ this results in the constraint \Rbpaper
\eq
|Y|^2 F(y_t) < 1.7~,
\eeq
where $y_t = m_t^2/\mhplus^2$ and
\eq
F(y) \equiv { y \over (y - 1)^2} \; [ y - 1 - \ln y ].
\eeq
This constraint is somewhat weaker than that from $b\to s\gamma$.

It is important to note that the phase of \bqbqbar\ mixing is unaffected by 
these new contributions. Also, the unitarity of the $3\times 3$ CKM matrix
holds in these models. Consequently, although the extraction of $|V_{td}|$
through the measurement of $\Delta M_d$ has to take the $H^+$ contribution
into account, the measurement of its phase in the asymmetry of
$\bdbarp\to\Psi\ks$ is unaffected by the presence of the new physics.

\ref\branco{G.C. Branco, \prl{44}{80}{504}; M. Gronau \etal,
\prd{37}{88}{250}.}

Particularly interesting are models of spontaneous CP violation, in which
the entire Lagrangian is CP invariant while the vacuum is not. (In the
Weinberg three-Higgs-doublet model \weingla, this possibility seems to have
already been ruled out by the experimental upper limit on the neutron
electric dipole moment \grosnir.) In this case CP is violated in (neutral
and charged) Higgs exchange, while natural flavor conservation leads to a
real CKM matrix \branco. Thus, the unitarity triangle becomes a straight
line, and the amplitude of \bqbqbar\ mixing is real.

\ref\fca{T.P. Cheng and M. Sher, \prd{35}{87}{3484}.}
\ref\fcb{A.S. Joshipura and S.D. Rindani, \plb{260}{91}{149}.}
\ref\bran{G.C. Branco, W. Grimus and L. Lavoura, \plb{380}{96}{119}.}
\ref\fcmodels{A. Antaramian, L.J. Hall and A. Rasin, \prl{69}{92}{1871};
Y-L. Wu and L. Wolfenstein, \prl{73}{94}{1762}; D. Atwood, L. Reina and A.
Soni, \prd{54}{96}{3296}.} 
\ref\hallwein{L. Hall and S. Weinberg, \prd{48}{93}{R979}.}

\noindent
{\it (ii) Flavor-Changing Neutral Scalars}

In the second class of models flavor-changing neutral scalar interactions
between quarks $i$ and $j$ exist, but are suppressed by factors $F_{ij}$
due to an approximate global symmetry. Such models have recently received
special attention \fca-\hallwein. In this case \bqbqbar\ mixing may also
receive large contributions from tree-level neutral Higgs exchange
amplitudes which carry new phases. Denoting the flavor-changing neutral
Higgs couplings by $(m_i/v)F_{ij}$ $(m_i>m_j,~v^{-2}=\sqrt{2}G_{\sss F}$),
their contributions to \bqbqbar\ mixing are given in the vacuum insertion
approximation by
\eq
M^{{\sss H}^0}_{12}(B_q) \approx {5\sqrt{2}\over 24} \,
{G_{\sss F} f^2_{B_q}m^3_{B_q}\over M^2_{{\sss H}^0}} \, F^2_{qb}~,
\eeq
where $M_{{\sss H}^0}$ includes possible complex mixing among several
neutral Higgs fields.

Comparing with mixing in the SM, we find
\eq
\label\neutHiggsmix
{M^{{\sss H}^0}_{12}(B_q) \over M^{\sss W}_{12}(B_q)} \approx 0.50 \,
\left({F_{qb} \over V_{tq}V^*_{tb}}\right)^2 \,
\left({100~{\rm GeV} \over M_{{\sss H}^0}}\right)^2~.
\eeq
Thus the neutral Higgs contributions may substantially modify the SM
prediction for \bqbqbar\ mixing. The unitarity triangle holds in this
model. However neither the magnitude of $V_{td}$ nor its phase can be
directly measured through $\Delta M_d$ and the asymmetry in
$\bdbarp\to\Psi\ks$, respectively. 

In models with specific predictions for $F_{ij}$, this leads to large
effects in an interesting range of Higgs masses. For instance,
neutral-Higgs contributions to \bqbqbar\ mixing are sizeable for Higgs
masses around 100 GeV when flavor-changing couplings have the form \fcb,
\bran
\eq
F_{qb}=V_{tq}V^*_{tb} ~~,
\eeq
and for Higgs masses of a few hundred GeV up to about a TeV in models in
which \fca
\eq
F_{qb}=\sqrt{m_q/m_b} ~~.
\eeq
In general, these new contributions carry unknown phases and have to be
added to the charged Higgs contributions [Eq.~\chHiggsratio].

In the presence of flavor-changing scalar interactions, the special case of
spontaneous CP violation does not, in general, forbid a phase in the CKM
matrix. For a particular choice of the softly-broken symmetry this phase
may, however, be very small \hallwein\ or may even vanish \bran. This would
imply that the unitarity triangle becomes a straight line, while the
\bqbqbar\ mixing amplitude carries a complex phase.

\subsection{Left-right symmetric models}

\ref\LRSmodels{R.N. Mohapatra and J.C. Pati, \prd{11}{75}{566},
{\bf D11} (1975) 2558; G. Senjanovic and R.N. Mohapatra,
\prd{12}{75}{1502}; G. Senjanovic, \npb{153}{79}{334}.}
\ref\heavyWR{G. Beall, M. Bander and A. Soni, \prl{48}{82}{848}.}

In left-right symmetric models \LRSmodels, the gauge group is extended to
$SU(2)_\lft \times SU(2)_\rht \times U(1)_{\sss B-L}$, along with a
discrete $L \leftrightarrow R$ symmetry. The right-handed CKM matrix is
then related to its left-handed counterpart: $V^\rht = V^\lft$ or $V^\rht
= (V^\lft)^*$. The right-handed $W_\rht$ can participate in weak processes
in the same way as the ordinary $W$ (although certain decays are forbidden
if the $\nu_\rht$ is too heavy). In particular, the $W_\rht$ can contribute
to \bqbqbar\ mixing through box diagrams, as in the SM. However, limits
from the $\kl$-$\ks$ mass difference constrain the $W_\rht$ to be heavier
than 1.4 TeV \heavyWR, which would render its effects in the $B$ system
negligible.

\ref\lightWR{F.I. Olness and M.E. Ebel, \prd{30}{84}{1034}; P. Langacker
and S.U. Sankar, \prd{40}{89}{1569}.}
\ref\lonwyler{D. London and D. Wyler, \plb{232}{89}{503}.}

If one abandons the discrete $L \leftrightarrow R$ symmetry, so that
$V^\rht$ is unrelated to $V^\lft$, the constraints from the $\kl$-$\ks$
mass difference can be evaded. For judicious choices of the form of
$V^\rht$, CP-conserving experimental data permit the $W_\rht$ to be
considerably lighter, $M_\rht \gsim 300$ GeV \lightWR. However, unless the
elements of $V^\rht$ are considerably fine-tuned, there will be large
contributions to the CP-violating parameter $\vert\epsilon_{\sss K}\vert$
\lonwyler. Assuming no such fine tuning, the $W_\rht$ is again constrained
to be heavy, $M_\rht \gsim 5$ TeV.

\ref\GroWak{M. Gronau and S. Wakaizumi, \prl{68}{92}{1814}.}
\ref\HouWyler{W.-S. Hou and D. Wyler, \plb{292}{92}{364}.}
\ref\BPsipi{J. Alexander \etal\ (CLEO Collaboration), \plb{341}{95}{435},
(E) {\it ibid.} {\bf B347}, 469 (1995).}

The above analysis assumes that $V^\lft$ has the same form as in the SM.
However, this need not be the case. For example, it was suggested \GroWak\
that $B$ decays might in fact be mediated by the $W_\rht$, instead of the
ordinary $W$. The long $B$ lifetime would then be interpreted as being due
to the heaviness of the $W_\rht$, rather than to the smallness of $V_{cb}$.
A variety of different forms for $V^\lft$ and $V^\rht$ were proposed
\GroWak, \HouWyler. In all cases, the $V^\rht_{cd}$ element was considerably
smaller than in the SM, leading to the prediction that $BR(b\to c{\bar c}d)
/ BR(b\to c{\bar c}s)$ is at most $O(10^{-4})$. However, the decay $B \to
\Psi \pi$ has since been observed with a branching ratio in agreement with
the SM \BPsipi, effectively ruling out all such models.

\ref\WakLRS{S. Wakaizumi, \plb{322}{94}{397}; T. Kurimoto, A. Tomita and S.
Wakaizumi, OU-HET-228, hep-ph/9512387.}

Our conclusion is therefore that there are no important new-physics effects
in the $B$ system within left-right symmetric models. The one possible
exception is if one considerably fine-tunes the right-handed CKM matrix
\WakLRS, but we do not consider such possibilities here.

\subsection{Supersymmetry}

\ref\SSM{For reviews, see \eg\ J. Bagger and J. Wess, {\it Supersymmetry
and Supergravity} (Princeton Univ.\ Press, Princeton, NJ, 1983); H.P.
Nilles, \prep{110}{84}{1}; P. Nath, R. Arnowitt and A.H. Chamseddine, {\it
Applied $N=1$ Supergravity} (World Scientific, Singapore, 1984); H.E. Haber
and G.L. Kane, \prep{117}{87}{1}; A.G. Lahanas and D.V. Nanopoulos,
\prep{145}{87}{1}.}

In the supersymmetric standard model (SSM) \SSM, the gauge group is
unchanged, but a plethora of new particles is added. These include the
supersymmetric partners of the SM particles, as well as a second Higgs
doublet (in some versions, additional Higgs representations are also
present). In the SSM there are thus a variety of new contributions to
\bqbqbar\ mixing. These come from box diagrams with internal (i) charged
Higgs bosons and charge 2/3 quarks, (ii) charginos and charge 2/3 squarks,
(iii) gluinos and charge $-1/3$ squarks, and (iv) neutralinos and charge
$-1/3$ squarks. The relative sizes of these new contributions, as well as
their phase information, depend on the version of the SSM.

\ref\softbreaking{L. Girardello and M. Grisaru, \npb{194}{82}{65}.}
\ref\BBMR{See, for example, S. Bertolini, F. Borzumati, A. Masiero and G.
Ridolfi, \npb{353}{91}{591}, and references therein.}
We first consider the minimal supersymmetric standard model (MSSM), which
is usually taken to mean the low-energy limit of the minimal
spontaneously-broken $N=1$ supergravity model. Here one typically imagines
that there is unification at some high scale ($M_{\sss X}$), and that
supersymmetry (SUSY) is broken at this scale by some unknown mechanism
(\eg\ a ``hidden sector'') which interacts only gravitationally with the
known fields. SUSY breaking is parametrized by soft breaking terms in the
supergravity lagrangian \softbreaking. The low-energy effects of SUSY
breaking, as well as the masses of the superpartners, are calculated by
running the renormalization group equations down from $M_{\sss X}$ to the
weak scale. The net effect is that, in addition to the usual gauge and
Yukawa couplings, the MSSM is described by only 4 new parameters. The
masses and mixings of all superpartners at low energy can be described in
terms of these 4 parameters. If one also requires that the spontaneous
breaking of $SU(2)_\lft \times U(1)_{\sss Y}$ be induced radiatively, there
is a further reduction in the number of SUSY parameters from 4 to 3 \BBMR.

\ref\SUSYFCNC{We follow here the analyses of T. Kurimoto,
\prd{39}{89}{3447}; S. Bertolini, F. Borzumati, A. Masiero and G. Ridolfi,
Ref.~\BBMR; G.C. Branco, G.C. Cho, Y. Kizukuri and N. Oshimo,
\plb{337}{94}{316}, \npb{44}{95}{483}; G. Couture and H. K\"onig,
\zpc{69}{96}{499}, UQAM-PHE-95-04, hep-ph/9511234.}
\ref\pdg{R.M. Barnett \etal\ (Particle Data Group), \prd{54}{96}{1}.}

Although there are several new SUSY contributions to \bqbqbar\ mixing, in
the MSSM these all have the same phase as in the SM, to a good
approximation. We consider them in turn \SUSYFCNC:

\topic{(i) charged Higgs bosons and charge 2/3 quarks} These contributions
have already been described above in the discussion of the
two-Higgs-doublet model (Sec.~{\it 2.3(i)}). Unless one goes to extremely
large values of $\tan\beta$, the charged Higgs couples the down-type quarks
only to the $t$-quark:
\eq
{\cal L}_{\sss H^\pm} \simeq {g\over \sqrt{2}} \cot\beta \, {m_t \over \mw}
\, H^+ \, {\bar t}_\rht V_{ti} d_{i\lft} ~.
\eeq
Thus the contribution to \bqbqbar\ mixing from charged Higgs bosons is
proportional to $(V_{tb}^* V_{tq})^2$, as in the SM.

\topic{(ii) charginos and charge 2/3 squarks} The couplings of the
$\wino^\pm$ and $\higgsino^\pm$ to up-type quarks are very similar to those
of the $W^\pm$ and $H^\pm$. (Note that the physical charginos are in
general linear combinations of $\wino^\pm$ and $\higgsino^\pm$.) In
particular, the $\wino^\pm$ couples only to left-handed up-type squarks:
\eq
{\cal L}_{\wino} = {g\over \sqrt{2}} \,
({\tilde u},~{\tilde c},~{\tilde t})_\lft V_{\sss CKM} {\overline{\wino}}
\, \gamma_\lft \left( \matrix{d \cr s \cr b \cr} \right),
\eeq
while in the limit of negligible down-type quark masses, the
$\higgsino^\pm$ couples mainly to right-handed squarks (assuming
non-extreme values of $\tan\beta$):
\eq
\label\higgsinocoup
{\cal L}_{\higgsino} \simeq {g\over \sqrt{2}} \, {1 \over \sin\beta} \,
{m_t \over \mw} \, {\tilde t}_\rht \, V_{ti} \, {\overline{\higgsino}}
\gamma_\lft d_i ~,
\eeq
where $\gamma_\lft = (1-\gamma_5)/2$. The contributions to \bqbqbar\ mixing
of both the $\wino^\pm$ and $\higgsino^\pm$ are proportional to $(V_{tb}^*
V_{tq})^2$, as in the SM. For the $\higgsino^\pm$ this follows directly
from the above equation, while in the case of the $\wino^\pm$, one uses the
unitarity of the CKM matrix, along with the fact that $m_{{\tilde u}_\lft}
= m_{{\tilde c}_\lft} \ne m_{{\tilde t}_\lft}$ in the MSSM, to arrive at
this result.

\topic{(iii) gluinos and charge $-1/3$ squarks} The important coupling of
the gluino (${\tilde g}$) to down-type quarks and squarks is
\eq
\label\gluinocoupling
{\cal L}_{\tilde g} = \sqrt{2} \, g_3 
({\tilde d},~{\tilde s},~{\tilde b})_\lft V_{\sss CKM} \, {\bar{\tilde g}} 
\, \gamma_\lft \left( \matrix{d \cr s \cr b \cr} \right). 
\eeq
(There is also a generation-diagonal coupling involving right-handed down
squarks, but this cannot contribute to \bqbqbar\ mixing.) Note that the
coupling is proportional to $V_{\sss CKM}$. This, along with the fact that
$m_{{\tilde d}_\lft} = m_{{\tilde s}_\lft} \ne m_{{\tilde b}_\lft}$ in the
MSSM, leads to a contribution to \bqbqbar\ mixing proportional to
$(V_{tb}^* V_{tq})^2$, as in the SM.

\topic{(iv) neutralinos and charge $-1/3$ squarks} The physical neutralinos
are linear combinations of the photino ($\photino$), the Zino ($\zino$),
and the two neutral Higgsinos (${\higgsino}_{1,2}^0$). The couplings of the
$\photino$ and $\zino$ to down-type quarks and squarks are similar to that
of the ${\tilde g}$ [Eq.~\gluinocoupling]:
\eq
\eqalign{
{\cal L}_{\photino} & = \left(-{1\over 3 } \, e\right) 
({\tilde d},~{\tilde s},~{\tilde b})_\lft V_{\sss CKM} \, {\bar\photino} 
\, \gamma_\lft \left( \matrix{d \cr s \cr b \cr} \right), \cr
{\cal L}_{\zino} & = {g\over\cos\theta_w} 
\left[ -{1\over 2} + {1\over 3} \sin^2 \theta_w \right]
({\tilde d},~{\tilde s},~{\tilde b})_\lft V_{\sss CKM} \, {\bar\zino} 
\, \gamma_\lft \left( \matrix{d \cr s \cr b \cr} \right). \cr}
\eeq
The dependence on $V_{\sss CKM}$ of the couplings of the ${\tilde\gamma}$
and ${\tilde Z}$ is just like that of the gluino, leading to a contribution
to \bqbqbar\ mixing which is proportional to $(V_{tb}^* V_{tq})^2$. As for
the neutral Higgsinos, their coupling to down-type quarks and squarks is
proportional to $M_{\sss D}/\mw\cos\beta$, which is negligible (unless one
goes to extremely large values of $\tan\beta$).

\endtopic

\noindent 
Thus, to a good approximation, in the MSSM all new SUSY contributions to
\bqbqbar\ mixing have the same phase as in the SM. Therefore the CP
asymmetries in $B$ decays will not be modified. The full expressions for
these contributions are quite complicated, so we do not reproduce them here
(we refer the reader to Ref.~\SUSYFCNC). The strongest constraints on the
SUSY parameters come from direct searches. For example, there are lower
bounds of 176 GeV and 45 GeV on squark and chargino masses, respectively
\pdg. The parameter space is sufficiently complicated that it is very
difficult to establish firm constraints from loop-level processes such as
$b\to s\gamma$. We note, however, that the effect of supersymmetry on
\bqbqbar\ can be quite significant -- for certain values of the parameters,
the total SUSY contribution to \bdbdbar\ mixing can be twice as large as
that of the SM.

\ref\falk{T. Falk and K. Olive, \plb{375}{96}{196}; S.A. Abel and J.M.
Fr\`ere, ULB-TH-96-15, hep-ph/9608251.}

Recently \falk\ it was suggested that CP violation could possibly come from
SUSY breaking alone, with a real CKM matrix. In this case, which is
essentially a superweak-type model, the unitarity triangle becomes a
straight line and all CP asymmetries in $B$ decays are very small.

We now turn to nonminimal SUSY models. For generic squark masses,
supersymmetric contributions enhance flavor-changing processes such as
$K^0$-${\overline{K^0}}$ well beyond their experimental values. Any
nonminimal SUSY model should address this problem (in the MSSM this
problem is resolved since, to a good approximation, the squarks are
degenerate).

\ref\alignment{Y. Nir and N. Seiberg, \plb{309}{93}{337}; M. Leurer,
Y. Nir and N. Seiberg, \npb{420}{94}{468}.}
\ref\CKN{A.G. Cohen, D.B. Kaplan and A.E. Nelson, \plb{388}{96}{588}.}
\ref\cohen{A.G. Cohen, D.B. Kaplan, F. Lepeintre and A.E. Nelson,
UW/PT-95/22, hep-ph/9610252.}

One class of nonminimal SUSY models solves the FCNC problem by imposing an
Abelian horizontal symmetry on the lagrangian \alignment. In this case the
quark mass matrices are approximately aligned with the squark mass-squared
matrices. This has the effect that the mixing matrix for
quark-squark-gluino couplings is close to a unit matrix, so that FCNC are
suppressed, even though the squarks may not be degenerate. In most such
models, the SUSY contributions to \bqbqbar\ mixing are quite small.
However, it is possible to construct models in which $M_{12}^{\sss
SUSY}(B_d^0)/M_{12}^{\sss SM}(B_d^0)$ is as large as 0.15, with a
negligible effect on \bsbsbar\ mixing. Since the phase of the SUSY
contribution is unknown, this can lead to measurable deviations from the SM
predictions in CP asymmetries involving $B_d^0$ decays. In another class of
nonminimal SUSY models, known as effective supersymmetry \CKN, the
suppression of FCNC's applies only to the first two families of squarks. In
this case new-physics effects on \bqbqbar\ mixing can be much larger
\cohen.

\ref\worah{M.P. Worah, \prd{54}{96}{2198}; N.G. Deshpande, B. Dutta, and S. 
Oh, \prl{77}{96}{4499}.}

Another approach which is often taken is to ignore the FCNC problem
altogether. One simply assumes that all SUSY parameters take the maximum
allowed values permitted by experiment. In such ``models'' one can have
non-negligible contributions to \bqbqbar\ mixing which have different
phases than in the SM. These contributions typically involve right-handed
squarks. For example \worah, the general coupling of a ${\higgsino}^\pm$ to
down-type quarks and up-type squarks can be written
\eq
{\cal L}_{\higgsino} \simeq {g\over \sqrt{2}} \, {1 \over \mw \sin\beta}
\, ({\tilde u},~{\tilde c},~{\tilde t})_\rht
{\tilde U}^{u^\dagger}_\rht U_\rht^u M_{\sss U} V_{\sss CKM}
{\overline{\higgsino}} \, \gamma_\lft \left( \matrix{d \cr s \cr b \cr}
\right),
\eeq
where ${\tilde U}^u_\rht$ ($U_\rht^u$) is the transformation matrix of the
right-handed up-type squarks (quarks) needed to diagonalize the squark
(quark) mass matrix. In the MSSM, ${\tilde U}^u_\rht = U_\rht^u$, leading
to a contribution to \bqbqbar\ mixing proportional to $(V_{tb}^* V_{tq})^2$
[Eq.~\higgsinocoup]. However, in general this relation need not hold, in
which case there can be new phases in this contribution to \bqbqbar\
mixing.

\ref\bigigabbiani{I.I. Bigi and F. Gabbiani, \npb{352}{91}{309}.}

As another example, consider again the contribution of gluinos and charge
$-1/3$ squarks to \bqbqbar\ mixing. In the MSSM there is no
intergenerational mixing among right-handed down-type squarks. As a
consequence, the contribution to \bqbqbar\ mixing of gluinos and charge
$-1/3$ squarks involves only left-handed squarks [Eq.~\gluinocoupling].
However, in nonminimal SUSY models, this need not be the case
\bigigabbiani\ -- there can be intergenerational mixing among right-handed
down-type squarks. In general, this mixing matrix is unrelated to $V_{\sss
CKM}$, so that there can be new phases in this contribution to \bqbqbar\
mixing.

The main problem with this approach is that there is little predictivity.
There are, in general, a very large number of parameters -- the masses of
the superpartners, their mixings, etc. Thus, although one can describe how
new phases can enter \bqbqbar\ mixing, it is virtually impossible to
analyse such effects in a systematic way.


\section{Penguin Decays and New Physics}

As discussed in the introduction, any new physics which contributes to
\bqbqbar\ mixing will also contribute to flavor-changing $B$ decays.
Before examining the new-physics contributions to such penguin decays, we
first review the SM predictions. Two aspects of these predictions are of
particular interest to us: (i) the actual size of the branching ratios for
various penguin decays, and (ii) the uncertainties, both experimental and
theoretical, of the predictions. New-physics effects will be considered
important in a particular penguin decay only if they change the branching
ratio by quite a bit more than the uncertainty in the SM prediction -- in
other words, we are looking for ``smoking gun'' signals of new physics in
such decays.

\subsection{The standard model}

\ref\buchalla{G. Buchalla and A.J. Buras, \npb{400}{93}{225}.}
\ref\Bqlllimit{F. Abe \etal\ (CDF Collaboration), \prl{76}{96}{4675}.}

\topic{$b\to q\gamma$, $q=d,s$} The lowest-order amplitude for the decay
$b\to q\gamma$ is \Alireview
\eq
\label\bsgammaamp
A(b\to q\gamma) = {G_{\sss F} \over \sqrt{2}} \, {e \over 2\pi} \, \sum_i
V_{ib}^* V_{iq} F_2(x_i) q^\mu \epsilon^\nu {\bar s} \sigma_{\mu\nu} (m_b
\gamma_\rht + m_s \gamma_\lft) b ~,
\eeq
where the sum is over the up-type quarks, and $q^\mu$ and $\epsilon^\mu$
are the photon's four-momentum and polarization, respectively. The function
$F_2$ is given by
\eq
F_2(x) = {x \over 24 (x-1)^2} \, \left[ 6 x ( 3 x - 2) \log x - (x-1) ( 8
x^2 + 5 x - 7 ) \right].
\eeq
Due to the smallness of the $u$- and $c$-quark masses, $F_2(x_u), F_2(x_c)
\ll F_2(x_t)$, so that the $b\to q\gamma$ amplitude is dominated by
$t$-quark exchange. A full quantitative treatment of these decays requires
the calculation of the important QCD corrections. Including these, the SM
branching ratios are \Alireview
\eq
\eqalign{
BR(B \to X_s \, \gamma) & = (3.2 \pm 0.58) \times 10^{-4} ~, \cr
BR(B \to X_d \, \gamma) & = (1.0 \pm 0.8) \times 10^{-5} ~. \cr}
\eeq
The uncertainties include both experimental errors [$m_t$, $B$ semileptonic
branching ratio] and theoretical errors [$\mu$ (the renormalization scale),
$\Lambda_{\sss QCD}$, and the ambiguity in the interpretation of $m_t$
(pole or running mass)], combined in quadrature. For $b\to d \, \gamma$,
$|V_{td}/V_{ts}| = 0.24 \pm 0.11$ has been used and combined in quadrature
with the other errors. The decay $b\to s\gamma$ has actually been measured
by the CLEO collaboration \CLEObsgamma:
\eq
BR(B \to X_s \, \gamma) = (2.32 \pm 0.67) \times 10^{-4} ~. 
\eeq
This measurement can be used to constrain models of new physics, as already
demonstrated in Eq.~\chHiggslimit .

\topic{$b\to q \, l^+ l^-$, $q=d,s$} This class of decays is rather
complicated theoretically \Alireview. First, one must calculate, at
next-to-leading order, the Wilson coefficients of ten local operators which
mix under renormalization. Second, one needs the matrix elements relevant
for $B \to X_{s,d} \, l^+ l^-$, which can be calculated using the spectator
model, along with $O(1/m_b^2)$ corrections. Finally, long-distance effects
due to $J/\Psi$ and $\Psi'$ resonances must also be taken into account. The
short-distance contribution for $b\to s \, l^+ l^-$, which can be measured
far away from the resonances, gives the following branching ratios, taken
from Ref.~\Alireview: 
\eq
\eqalign{
BR(B \to X_s \, e^+ e^-) & = (8.4 \pm 2.2) \times 10^{-6} ~, \cr
BR(B \to X_s \, \mu^+ \mu^-) & = (5.7 \pm 1.3) \times 10^{-6} ~, \cr
BR(B \to X_s \, \tau^+ \tau^-) & = (2.6 \pm 0.5) \times 10^{-7} ~. \cr}
\eeq
For $b\to d \, l^+ l^-$, the branching ratios are
\eq
\eqalign{
BR(B \to X_d \, e^+ e^-) & = (4.9 \pm 4.3) \times 10^{-7} ~, \cr
BR(B \to X_d \, \mu^+ \mu^-) & = (3.3 \pm 2.8) \times 10^{-7} ~, \cr
BR(B \to X_d \, \tau^+ \tau^-) & = (1.5 \pm 1.3) \times 10^{-8} ~. \cr}
\eeq
For all decays, the errors come from the same sources as in $b\to q\gamma$:
$m_t$, $B$ semileptonic branching ratio, the renormalization scale $\mu$,
$\Lambda_{\sss QCD}$, and $|V_{td}/V_{ts}| = 0.24 \pm 0.11$.

\topic{$B^0_q \to l^+ l^-$, $q=d,s$} This decay can be calculated quite
precisely in the SM. By including the QCD corrections \buchalla, the
renormalization-scale uncertainty is reduced to $O(1\%)$. There is still
some hadronic uncertainty, parametrized by the $B$-meson decay constant
$f_{\sss B}$. The branching ratios are \Alireview
\eq
\eqalign{
BR(B^0_s \to \tau^+ \tau^-) & = (7.4 \pm 2.1) \times 10^{-7} 
~(f_{\sss B_s}/232~{\rm MeV})^2 ~, \cr
BR(B^0_s \to \mu^+ \mu^-) & = (3.5 \pm 1.0) \times 10^{-9} 
~(f_{\sss B_s}/232~{\rm MeV})^2 ~, \cr
BR(B^0_d \to \tau^+ \tau^-) & = (3.1 \pm 2.9) \times 10^{-8} 
~(f_{\sss B_d}/200~{\rm MeV})^2 ~, \cr
BR(B^0_d \to \mu^+ \mu^-) & = (1.5 \pm 1.4) \times 10^{-10} 
~(f_{\sss B_d}/200~{\rm MeV})^2 ~. \cr}
\eeq
(The branching ratios to $e^+e^-$ are some 5 orders of magnitude smaller
than those for $\mu^+\mu^-$.) The error in the $B_s^0$ branching ratios is
due to the uncertainty, both experimental and theoretical, in the top-quark
mass. The $B_d^0$ branching ratios have a larger error due to the $V_{td}$
CKM matrix element: $|V_{td}/V_{ts}| = 0.24 \pm 0.11$. At present, the best
upper limits are $BR(B^0_s \to \mu^+ \mu^-) < 8.4 \times 10^{-6}$ and
$BR(B^0_d \to \mu^+ \mu^-) < 1.6 \times 10^{-6}$ \Bqlllimit, with no
significant limits on the $\tau^+\tau^-$ final state. 

\ref\buras{A.J. Buras, M. Jamin and M.E. Lautenbacher, \npb{408}{93}{209};
M. Ciuchini, E. Franco, G. Martinelli and L. Reina, \npb{415}{94}{403}.}
\ref\DeshTram{See, for example, N.G. Deshpande and J. Trampetic,
\prd{41}{90}{895}.}
\ref\grorosner{M. Gronau and J.L. Rosner, \plb{376}{96}{205}.}

\topic{Gluon-mediated exclusive hadronic decays} These arise from the
quark-level process $b\to q q'{\bar q}'$. Throughout this paper we will
refer to such loop-level decays as ``hadronic penguins.'' There are two
ingredients needed to calculate the rates for hadronic penguin decays in
the SM. First, the rates for the quark-level decays $b\to sq{\bar q}$ and
$b\to dq{\bar q}$ are computed. This is done similarly to the decays $b\to
s \, l^+ l^-$ and $b\to d \, l^+ l^-$: the Wilson coefficients of a variety
of operators are calculated as one renormalizes down from the weak scale to
the $b$-mass \buras. This can be done with reasonable precision. Second,
one calculates the hadronic matrix elements for the hadronization of the
final-state quarks into particular final states \DeshTram. It is this step
which introduces enormous uncertainty. These hadronic matrix elements are
typically evaluated using the factorization approximation. Unfortunately,
it is difficult to estimate the error incurred by applying this
approximation to penguin decays. The predicted rates for exclusive hadronic
penguin decays can easily be in error by a factor of 2 to 3. (Much of this
uncertainty cancels in the ratio of rates of corresponding $b \to d$ and
$b\to s$ processes, which is given in the SM by $|V_{td}/V_{ts}|^2$
\grorosner.)

Since the SM predictions for hadronic penguins have considerable
uncertainties, if one wants an unmistakable signal of physics beyond the
SM in such decays, the new-physics effects must be {\it enormous} -- they
must change the SM rates by an order of magnitude or more. It is therefore
sufficient for our purposes to obtain approximate, order-of-magnitude
estimates for both the SM and new-physics effects. To this end, we will use
the following approximate form for the amplitude of the SM gluonic penguin
contribution to the decay $b \to q q' {\bar q}'$:
\eq
\label\hadpenguin
A_{penguin} \sim {\alpha_s(m_b) \over 12\pi} \, \log\left({m_t^2\over
m_b^2}\right) \, V_{tq} V_{tb}^* \simeq 0.04 \, V_{tq} V_{tb}^* ~.
\eeq
(Note that the coefficient $0.04$ is about the same size as the largest of
the Wilson coefficients of penguin operators \buras .) 

\ref\CLEOpeng{D.M. Asner \etal\ (CLEO Collaboration), \prd{53}{96}{1039}.}

This expression can be used to estimate the order-of-magnitude rates for 
$b \to s$ and $b \to d$ penguins in the SM. For example, the branching
ratio for $B_d^0 \to \pi^+\pi^-$, which is dominated by the tree-level
$b\to u{\bar u}d$ amplitude, is $O(10^{-5})$. Comparing this decay with
$b\to s$ penguins, which dominate $B_d^0 \to \pi^- K^+$, we find
\eq
\left\vert {A_{penguin}(b\to s) \over A_{tree}(b\to u{\bar u}d)}
\right\vert \sim \left\vert {0.04 \, V_{tb}^* V_{ts} \over V_{ub}^* V_{ud}}
\right\vert \sim 0.5~,
\eeq
where we have used $|V_{ts}| \simeq |V_{cb}|$ and $|V_{ub}/V_{cb}| = 0.08$.
This is consistent with the observation of a combined sample of
$B_d^0\to\pi^+\pi^-$ and $B_d^0\to\pi^- K^+$ decays \CLEOpeng, in which
about equal mixtures of both modes are most likely. Thus, assuming that the
hadronic matrix elements of tree and penguin operators have similar
magnitudes, we expect pure $b\to s$ penguin hadronic decays (\eg\
$B^+\to\pi^+\ks$, $B^0_d \to \phi\ks$) to have branching ratios of
$O(10^{-5})$. $b\to d$ penguins can be analyzed similarly:
\eq
\label\btodpenguin
\left\vert {A_{penguin}(b\to d) \over A_{tree}(b\to u{\bar u}d)}
\right\vert \sim \left\vert {0.04 \, V_{tb}^* V_{td} \over V_{ub}^* V_{ud}}
\right\vert ~.
\eeq
Since $1.4 \le |V_{td}/V_{ub}| \le 4.6$ \alilon, the penguin amplitude is
about 1/10 the size of the tree amplitude, and may be larger if the
hadronic penguin matrix elements are enhanced relative to those of tree
amplitudes. Thus, pure $b\to d$ penguin hadronic decays (\eg\ $B^+\to
K^+\ks$) should have branching ratios of $O(10^{-7})$, or somewhat larger.

As an aside, we note that Eq.~\btodpenguin\ demonstrates why penguin
contamination is a concern in the extraction of $\sin 2\alpha$ using the CP
asymmetry in $\bdbarp\to\pi^+\pi^-$. According to this estimate, the
penguin amplitude can be as much as $\sim 15\%$ of the tree amplitude in
magnitude (or even larger, if penguin matrix elements are enhanced), and
with a different phase. This can lead to considerable uncertainty in the
extraction of $\sin 2\alpha$, and shows why isospin techniques \isospin\
are necessary to remove the penguin contamination.

\ref\bsw{M. Bauer, B.Stech and M. Wirbel, \zpc{34}{87}{103}.}
\ref\Alam{M.S. Alam \etal\ (CLEO Collaboration), \prd{50}{94}{43}.}
\ref\fleischer{R. Fleischer, \zpc{62}{94}{81}; N.G. Deshpande and X.-G.
He, \plb{336}{94}{471}.}

\topic{Electroweak-penguin-dominated exclusive hadronic decays} Here, the
diagrams contributing to the quark-level process $b\to q q'{\bar q}'$ 
consist of $\gamma$ and $Z$ penguins and box diagrams. Of these, the
diagram involving $Z$ exchange is the most important since it is enhanced
by a factor of $m_t^2/\mw^2$. Throughout this paper we will refer to such
processes as ``electroweak penguins.'' As with hadronic penguins, the
calculation of SM rates for exclusive electroweak penguin decays suffers
from large uncertainties in the hadronic matrix elements. We will therefore
again use an approximate form for the amplitude of the SM electroweak
penguin contribution to the decay $b \to q q' {\bar q}'$. This can be
obtained from Eq.~\hadpenguin\ through the replacements $\alpha_s(m_b) \to
\alpha_2(m_b)$ and $\log\left(m_t^2/m_b^2\right) \to m_t^2/\mw^2$. There is
an additional factor of 2 due to a larger Wilson coefficient for the
electroweak penguin operator. Therefore,
\eq
\label\EWpenguin
A_{\sss EWP} \sim {\alpha_2(m_b) \over 6\pi} \, \left({m_t^2\over \mw^2}
\right) \, V_{tq} V_{tb}^* \simeq 0.008 \, V_{tq} V_{tb}^* ~.
\eeq
Note that, if one combines the final-state $q$ and ${\bar q}'$ quarks to
form a meson, there is an additional color-suppression factor, $a_2$ \bsw.
Compared to color-allowed decays (\ie\ forming a meson from $q'$ and
${\bar q}'$), which are parametrized by $a_1$, this suppression is $a_2/a_1
= 0.2$ \Alam.

Comparing Eqs.~\hadpenguin\ and \EWpenguin, we note that electroweak
penguin amplitudes are suppressed relative to their hadronic penguin
counterparts by a factor 0.2. Therefore we expect $b\to s$ and $b \to d$
electroweak penguin decays to have branching ratios in the range
$10^{-7}$-$10^{-6}$ and $10^{-9}$-$10^{-8}$, respectively \fleischer. Some
examples of decays which are dominated by electroweak penguins are $B^0_s
\to \phi\pi^0$ ($b \to s$) and $B^+ \to \phi\pi^+$ ($b \to d$).

\endtopic
\noindent
At this point several observations are in order. From the above summary, we
see that the SM $b\to d$ penguins have branching ratios which are about 
one to two orders of magnitudes smaller than their $b\to s$ counterparts.
Therefore, unless the new physics has a large influence on the $b\to d$
FCNC, its effects on $b\to s$ penguins will be detected first. On the other
hand, CP asymmetries involving $B_d^0$ decays are likely to be
measured---and hence will reveal the presence of new physics---well before
those involving $B_s^0$ mesons. So, from a practical point of view, this
poses a bit of a problem. That is, even if new physics is detected in CP
violation in $B_d^0$ decays, it will be possible to test its nature by
looking at $b\to d$ penguin decays only if the small SM branching ratios
for these processes are significantly enhanced. Conversely, if one finds
new-physics effects in $b\to s$ penguins, it will be difficult to determine
its origin by looking at CP asymmetries in $B_s^0$ decays.

This having been said, however, the situation is not quite so bleak.
Although it is possible to construct models of physics beyond the SM in
which only one of the $b\to d$ or $b\to s$ FCNC's is changed, in practice
both FCNC's are affected in most new-physics models. For example, in models
with $Z$-mediated FCNC's, the $b\to d$ or $b\to s$ FCNC's are indeed
described by different parameters. However, both of these FCNC's arise due
to the mixing of the charge $-1/3$ quarks with an exotic vector singlet
quark. It would be difficult to imagine that only one of the two FCNC's is
induced, although this is a logical possibility. As another example,
consider models with four quark generations. The enlarging of the CKM
matrix to $4\times 4$ will, in general affect both FCNC's. Thus, in looking
for new physics, the measurements of CP asymmetries and $B$ penguins are
complementary, and it is quite likely that, should new physics be
discovered, its nature will be revealed only by studying both $B$
asymmetries and decays.

We now turn to the new-physics contributions to the above penguin decays.

\subsection{Four generations} 

Models with four generations have a number of new parameters which can
enter in $B$ decays: $m_{t'}$ and the CKM matrix elements involving the
$t'$ quark. Furthermore, the strongest constraints on $V_{ts}$ and $V_{td}$
in the SM come from the unitarity of the $3 \times 3$ CKM matrix. When this
matrix is enlarged to be $4 \times 4$, these constraints are weakened, so
that, in effect, $V_{td}$ and $V_{ts}$ are also unknown parameters. 

The parameter space of four-generation models is therefore quite large.
Rather than exploring the entire space, we will simply present an
``existence proof." That is, we will show that it is possible to choose
values of $m_{t'}$ and the CKM matrix elements, consistent with
experimental data, which significantly affect both CP asymmetries and $B$
penguin decays.

As discussed in Sec.~2.1, the experimental value of \bdbdbar\ mixing can be
reproduced if $V_{td} \sim 0$, $V_{t'd} = 0.005$, $V_{tb} = V_{t'b} \simeq
1/\sqrt{2}$, and $m_{t'} = 480$ GeV. The phase of this mixing may, however,
be quite different than in the SM. In this scenario, all penguin decays
involving the $b$-$d$ FCNC will also be dominated by the fourth generation.
Below we consider the effects of this choice of parameters on $b\to d$
penguins.

We first consider the decays $b\to q\gamma$. The experimental measurement
of $b\to s\gamma$ can be easily accommodated by adjusting the $t$ and $t'$
contributions. However, the decay $b\to d\gamma$ will involve only $t'$
exchange. Using Eq.~\bsgammaamp, we find that $|V_{t'd} V_{t'b}^* \,
F_2(x_{t'})| = 0.001$, as compared to the SM value of $|V_{td} V_{tb}^* \,
F_2(x_{t})| = 0.002$. Therefore, this choice of parameters will result in a
branching ratio for $b\to d\gamma$ which is roughly 4 times smaller than in
the SM. Given the large uncertainties in the SM prediction, this cannot be
considered an unmistakable signal of new physics.

Now consider hadronic penguins. From Eq.~\hadpenguin, we have
\eq
\left\vert {A_{penguin}^{4-gen} \over A_{penguin}^{\sss SM}} \right\vert = 
\left\vert {\log\left( m_{t'}^2/m_b^2 \right) \, V_{t'd} V_{t'b}^* \over
\log\left( m_t^2/m_b^2 \right) \, V_{td} V_{tb}^*} \right\vert = 0.46 ~.
\eeq
Thus, in this model, the branching ratios for exclusive $b\to d$ hadronic
penguins will be a factor of $\sim 5$ smaller than in the SM. Given the
uncertainties in the SM predictions this is also a marginal ``smoking gun''
signal of new physics.

The situation is better for electroweak penguins. From Eq.~\EWpenguin,
\eq
\left\vert {A_{\sss EWP}^{4-gen} \over A_{\sss EWP}^{\sss SM}} \right\vert
= \left\vert {\left( m_{t'}^2/\mw^2 \right) \, V_{t'd} V_{t'b}^* \over
\left( m_t^2/\mw^2 \right) \, V_{td} V_{tb}^*} \right\vert = 2.8 ~.
\eeq
The branching ratios for exclusive $b\to d$ electroweak penguins are thus a
factor of $\sim 8$ larger than in the SM. This same enhancement applies to
the decays $B_d^0 \to l^+l^-$. In both cases, this would be a quite
convincing signal of physics beyond the SM. 

Thus, in models with four generations, we have shown that there are regions
of parameter space in which both $B$ CP asymmetries and $B$ penguin decays
are significantly affected. If a discrepancy with the SM is found in the
measurement of the CP asymmetries, the study of the decays can help pin
down the new-physics parameters. Admittedly, in the particular example we
have chosen, the branching ratios for the affected processes are all small,
of $O(10^{-7})$ or smaller. But the key point here is that the various
penguin processes depend differently on the masses of the $t$ and $t'$
quarks. It is thus straightforward to find a set of parameters, consistent
with current experimental data, in which the $b\to s$ FCNC decays are
affected. In this case it would be CP asymmetries involving $B_s^0$ decays
which would be altered. Of course, since the four-generation CKM matrix is
$4\times 4$, in the general case both the $b\to d$ and $b\to s$ FCNC's will
be changed from the SM, affecting {\it all} CP asymmetries and penguin
decays.

\subsection{$Z$-mediated FCNC's}

For the process $b\to qf{\bar f}$, the amplitude due to $Z$-mediated FCNC's
is 
\eq
{\cal M}_{\sss Z-FCNC} = {4 G_{\sss F} \over \sqrt{2}} \, U_{qb} \,
{\bar q} \gamma^\mu \gamma_\lft b \, {\bar f} \left[ g_\lft^f \gamma_\mu
\gamma_\lft + g_\rht^f \gamma_\mu \gamma_\rht \right] f ~.
\eeq
The rate for $b\to qf{\bar f}$ is simply
\eq
\Gamma(b\to qf{\bar f}) = {G_{\sss F}^2 m_b^5 \over 192 \pi^3} \,
|U_{qb}|^2 \left[ \left( g_\lft^f \right)^2 + \left( g_\rht^f \right)^2
\right],
\eeq
while the rate for $B_q^0 \to l^+ l^-$ is
\eq
\label\FCNCBllrate
\Gamma(B_q^0 \to l^+ l^-) = {G_{\sss F}^2 \over 16 \pi} \, \tau_{\sss B_q}
f_{\sss B_q}^2 M_{\sss B_q} m_\lft^2 |U_{qb}|^2 ~.
\eeq
With these expressions in hand, we can now calculate or estimate the
contributions to penguin decays due to $Z$-mediated FCNC's. In all cases,
when presenting numbers, we use the upper limit $U_{qb} < 0.0017$
[Eq.~\FCNClimit].

The upper limit of $U_{qb} < 0.0017$ is in fact derived from the
experimental limit on the branching ratio of $B \to \mu^+\mu^- X$ [see
Sec.~2.2]. Thus, with this value of $U_{qb}$, $Z$-mediated FCNC models
``predict'' that $BR(B \to \mu^+\mu^- X) = 5 \times 10^{-5}$. For $b\to s$
FCNC's, this is roughly an order of magnitude larger than the SM
prediction, while for $b\to d$ transitions it is about 2 orders of
magnitude larger. If the branching ratios for the decays $B\to X_s \, l^+
l^-$ and $B\to X_d \, l^+ l^-$ are observed to be consistent with the SM
predictions, this will imply that $|U_{sb}| \lsim 6 \times 10^{-4}$ and
$|U_{db}| \lsim 1 \times 10^{-4}$. In both cases, this implies that the
new-physics effects in \bqbqbar\ mixing are negligible. Conversely, if
\bqbqbar\ mixing is significantly affected by $Z$-mediated FCNC's, then one
expects to see a substantial enhancement of the branching ratios for $B\to
X_s \, l^+ l^-$ and/or $B\to X_d \, l^+ l^-$.

We now turn to $B^0_q \to l^+ l^-$, $q=d,s$. From Eq.~\FCNCBllrate, the
rates for these processes, due only to $Z$-mediated FCNC's, are
\eq
\eqalign{
BR(B^0_s \to \tau^+ \tau^-)|_{\sss Z-FCNC} & < 1.6 \times 10^{-5} 
~(f_{\sss B_s}/232~{\rm MeV})^2 ~, \cr 
BR(B^0_s \to \mu^+ \mu^-)|_{\sss Z-FCNC} & < 5.8 \times 10^{-8} 
~(f_{\sss B_s}/232~{\rm MeV})^2 ~, \cr
BR(B^0_d \to \tau^+ \tau^-)|_{\sss Z-FCNC} & < 1.2 \times 10^{-5} 
~(f_{\sss B_d}/200~{\rm MeV})^2 ~, \cr
BR(B^0_d \to \mu^+ \mu^-)|_{\sss Z-FCNC} & < 4.2 \times 10^{-8} 
~(f_{\sss B_d}/200~{\rm MeV})^2 ~. \cr}
\eeq
Thus, if the FCNC parameters $U_{qb}$ have values near the present upper
limits, the predicted rates for $B_s^0 \to l^+ l^-$ and $B_d^0 \to l^+ l^-$ 
are respectively about 20 and 300-400 times larger than those expected in
the SM.

Turning to hadronic and electroweak penguins, we note that there are three
types of comparisons which can be made. First consider decays such as
$B_s^0 \to \ks\ks$ ($b\to s$) or $B_d^0 \to \ks\ks$ ($b\to d$), which
receive contributions from ordinary (gluonic) penguins and
color-suppressed $Z$-mediated FCNC's. For these decays we have
\eq
\left\vert {A_{\sss Z-FCNC} \over A_{\sss SM}} \right\vert \sim \left\vert
{a_2 \over a_1} \, {U_{qb} \sqrt{\left(g_\lft^d\right)^2 +
\left(g_\rht^d\right)^2} \over 0.04 \, V_{tb}^* V_{tq}} \right\vert = 2.2
\, \left\vert {U_{qb}\over V_{tq}} \right\vert < 
\cases{0.1, & $q=s$,\cr 0.4, & $q=d$,\cr}
\eeq
where we have taken $|U_{qb}/V_{cb}| < 0.044$ and $|V_{td}/V_{cb}| \simeq
0.24$ in our estimates of the ratio. The branching ratios for this type of
$B$ decays will therefore not be significantly affected by $Z$-mediated
FCNC's. 

Next we have decays such as $B_d^0 \to \phi\ks$ ($b\to s$) or $B_s^0 \to
\phi\ks$ ($b\to d$). Here there are contributions from ordinary gluonic
penguins and color-allowed $Z$-mediated FCNC's. Then
\eq
\left\vert {A_{\sss Z-FCNC} \over A_{\sss SM}} \right\vert \sim \left\vert
{U_{qb} \, g_{\sss V}^s \over 0.04 \, V_{tb}^* V_{tq}} \right\vert = 8.8\,
\left\vert {U_{qb}\over V_{tq}} \right\vert < 
\cases{0.4, & $q=s$,\cr 1.6, & $q=d$.\cr}
\eeq
In this case, $Z$-mediated FCNC's will not much affect the $b\to s$ penguin
decays, but the branching ratios for $b\to d$ penguins can be increased by
a factor of 3-4. Given the uncertainties in the SM predictions for such
decays, this cannot be considered significant. However, if the value of
$V_{td}$ is in fact smaller than 0.24 (in this model it can be as small as
0.07), then the branching ratios for $b \to d$ penguins will be
correspondingly increased. We consider this to be a marginal prediction,
since it depends sensitively on the true value of $V_{td}$.

Finally, decays such as $B_s^0 \to \phi\pi^0$ ($b\to s$) or $B^+ \to
\phi\pi^+$ ($b\to d$) have contributions from ordinary electroweak penguins
(but not gluonic penguins) and $Z$-mediated FCNC's. Here
\eq
\eqalign{
\left\vert {A_{\sss Z-FCNC} \over A_{\sss SM}} \, (b\to s) \right\vert &
\sim \left\vert {U_{sb} 
\over 0.008 \,
V_{tb}^* V_{ts}} \right\vert < 5.5 ~, \cr 
\left\vert {A_{\sss Z-FCNC} \over A_{\sss SM}} \, (b\to d) \right\vert &
\sim \left\vert {U_{db} 
\over 0.008 \, V_{tb}^* V_{td}}
\right\vert < 22.9 ~. \cr} 
\eeq
The effects of $Z$-mediated FCNC's on such decays are clearly enormous. The
branching ratios for pure electroweak penguin decays can be increased by as
much as a factor of $\sim 25$ ($b\to s$) or $\sim 500$ ($b\to d$)! Clearly
this is a ``smoking gun'' signal of new physics.

On the topic of hadronic penguin decays, there is another possibility which
should be mentioned. In the SM, the decay $B^+ \to \pi^+\pi^0$ occurs
principally via a tree-level amplitude -- there is no gluonic penguin and
the electroweak penguin is much suppressed. One therefore does not expect
to find CP violation in this mode. However, $Z$-mediated FCNC's also
contribute to this decay. Comparing this new-physics contribution with the
SM tree-level amplitude, we find
\eq
\left\vert {A_{\sss Z-FCNC} \over A_{\sss SM}} \right\vert \sim \left\vert
{U_{db} \, (g_{\sss A}^u - g_{\sss A}^d) \over V_{ub}^* V_{ud}} \right\vert
\lsim 0.5 ~. 
\eeq
The $Z$-mediated FCNC contribution to this decay could therefore be
substantial. If there is a significant strong phase difference between the
two amplitudes, following for instance from different rescattering in the
$I=2$ channel, this could lead to direct CP violation in this decay mode.
If found, this would be another clear signal of this particular type of new
physics.

\ref\handoko{L.T. Handoko and T. Morozumi, \mpla{10}{95}{309}; (E) {\it
ibid.} {\bf A10}, 1733 (1995).} 

$Z$-mediated FCNC's can also contribute to the decays $b\to q\gamma$ at the
one-loop level. However, the calculations are highly model-dependent, since
it is necessary to include the new vector-singlet quark(s) with which the
SM charge $-1/3$ quarks mix. The authors of Ref.~\handoko\ considered the
case of a single vector-singlet quark, and included the contribution of the
Higgs boson as well. For $|U_{qb}| < 0.0017$ they found that the
contribution to $b\to s\gamma$ was unimportant, but that the branching
ratio for $b\to d\gamma$ could be changed significantly, depending on the
values of the masses of the exotic quark and the Higgs boson. Since there
is quite a bit of model dependence, we do not consider this to be a clean
signal of new physics.

In summary, if $Z$-mediated FCNC's contribute significantly to \bqbqbar\
mixing, they will also lead to large effects in a variety of penguin
decays. The present experimental upper limit on the FCNC parameters is
$|U_{qb}| < 0.0017$ ($q=d,s$). This value for the parameters leads to
unmistakable effects in $b\to q \, l^+ l^-$, $B^0_q \to l^+ l^-$, 
and electroweak penguins. In addition,
$Z$-mediated FCNC's may lead to direct CP violation in decay modes such as
$B^+ \to \pi^+\pi^0$.

\subsection{Multi-Higgs-doublet models}

\noindent
{\it (i) Natural Flavor Conservation}

In these models the new contributions to rare $B$ decays come about through
amplitudes in which charged-Higgs exchange replaces the SM $W$ exchange.
As noted in Eq.~\chHiggslimit\ and the surrounding discussion, the
measurement of $b \to s\gamma$ already excludes a region of parameter
space with low $\mhplus$ and high $|Y|$.

\ref\savage{J.L. Hewett, S. Nandi and T.G. Rizzo, \prd{39}{89}{250}; M.
Savage, \plb{266}{91}{135}.} 
\ref\skiba{Sizeable neutral-Higgs contributions to $B^0_s \to \tau^+
\tau^-$ occur only for very large values of $X$, see W. Skiba and J.
Kalinowski, \npb{404}{93}{3}.}

In spite of this, charged-Higgs exchange may have significant effects on
other rare $B$ decays. The processes $B^0_q \to l^+ l^-$, $q=d,s$ were
studied in multi-Higgs models in Ref.~\savage. Neglecting small
contributions from neutral pseudoscalar Higgs (which are proportional to
$m^2_b/M^2_{\sss P}$, where $M_{\sss P}$ is the pseudoscalar mass) \skiba,
one finds the following expression for the ratio of charged-Higgs and SM
amplitudes:
\eq
\label\llratio
{A_{{\sss H}^+}(B^0_q \to l^+ l^-) \over A_{\sss SM}(B^0_q \to l^+ l^-)} =
{{1\over 2} x_t B(y_t)|Y|^2\over B(x_t) - C(x_t)} \approx -2.25 \, B(y_t)
\, |Y|^2~,
\eeq
where 
\eq
B(x) = {1\over 4}\left[{x\over 1-x} + {x\over (1-x)^2}\ln x\right]
~,~~~~~~~ 
C(x) = {x\over 4}\left[{3-{1\over 2}x\over 1-x} + 
{{3\over 2}x+1\over(1-x)^2}\ln x\right]~. 
\eeq
Since $B(y_t)$ is negative, the two amplitudes add up constructively. 

To illustrate the effect, let us consider the case $\mhplus = 400$ GeV,
$Y=3$ discussed in Sec.~{\it 2.3(i)}. For these parameters, the ratio of
Eq.~\llratio\ becomes 2.5. Thus, the $B^0_q \to l^+ l^-$ rates are expected
to be an order of magnitude larger than in the SM, while charged-Higgs
exchange dominates \bqbqbar\ mixing.

\ref\bqllMHDM{W.S. Hou and R.S. Willey, \plb{202}{88}{591},
\npb{326}{89}{54}; B. Grinstein, M.J. Savage and M.B. Wise,
\npb{319}{89}{271}; N.G. Deshpande, K. Panose and J. Trampetic,
\plb{308}{93}{322}.}

The decays $b \to q \, l^+ l^-$ in multi-Higgs-doublet models are more
complicated than $B^0_q \to l^+ l^-$, since there are more operators which
can contribute to this process. We refer to Ref.~\bqllMHDM\ for the details
of the computation, but the conclusions are as follows. For those values of
$|Y|$ and $\mhplus$ for which \bqbqbar\ mixing is dominated by
charged-Higgs exchange (\ie\ large $|Y|$), the decays $b \to q \, l^+ l^-$
can be enhanced by a factor of about 2. For $b \to d \, l^+ l^-$ this is
within the error of the SM prediction, but it is a significant effect for 
$b \to s \, l^+ l^-$.

\ref\dongol{J.F. Donoghue and E. Golowich, \prd{37}{88}{2542}; A.J. Davies, 
T. Hayashi, M. Matsuda and M. Tanimoto, \prd{49}{94}{5882}; N.G. Deshpande 
and X-G. He, \plb{336}{94}{471}; C.Q. Geng, P. Turcotte and W.S. Hou,
\plb{339}{94}{317}.}

Finally, we turn to hadronic penguin decays. Model-dependent studies 
\dongol\ show that, once the constraint from $b \to s\gamma$ is taken into 
account, the effect of charged-Higgs exchange can change the SM predictions
by no more than a few tens of percent. Since these predictions suffer from
large hadronic uncertainties, the rates of these processes cannot signal
charged-Higgs effects. 

\noindent
{\it (ii) Flavor Changing Neutral Scalars}

In models without NFC, one can have flavor-changing processes mediated by
neutral scalars (FCNS). The flavor-changing couplings between quarks of
flavor $i$ and $j$ are parametrized as
$(m_i/v)F_{ij}(\gamma_5/2)~(m_i>m_j)$, while the flavor-conserving
couplings are $m_f/v$. Thus there are also contributions to penguin decays
due to neutral scalar exchange. Consider the decay $b \to q f {\bar f}$,
where $f$ can be a quark or a lepton. The rate for this decay, due to FCNS
alone, is
\eq
\Gamma(b\to q f {\bar f})|_{\sss FCNS} = { G_{\sss F}^2 m_b^5 \over 3072
\pi^3} \left({m_b m_f \over M_{\sss H^0}^2}\right)^2 \left| F_{qb}
\right|^2~.
\eeq
This rate is clearly maximized when the fermion $f$ is as massive as
possible. Consider then the decay $b\to q \, \tau^+\tau^-$. For $M_{\sss
H^0} = 100$ GeV, $F_{db} = 0.02$ is essentially the largest value possible
due to constraints from \bdbdbar\ mixing. On the other hand, $F_{sb}$ has
no similar constraint, so we take $F_{sb} = \sqrt{m_s/m_b} = 0.17$. Then
\eq
BR(b \to q \, \tau^+ \tau^-)|_{\sss FCNS} = 
\cases{2.4 \times 10^{-7}, & $q=s$,\cr 4.3\times 10^{-9}, & $q=d$.\cr}
\eeq
The branching ratio for $b \to d \tau^+ \tau^-$ is only about 1/3 of the SM
expectation, well within the errors of the prediction. This decay can
therefore not be used to find effects of neutral scalars. On the other
hand, the decay $b \to s \, \tau^+ \tau^-$ could be significantly affected
by flavor-changing neutral scalars, since its new-physics branching ratio
is of the same order as the SM prediction. However, note that we have
selected almost maximal values for the new-physics parameters. If $F_{sb}$
is smaller, or $M_{\sss H^0}$ larger, than the values we have chosen, the
new-physics contribution to $b \to s \, \tau^+ \tau^-$ would then diminish
considerably, although there could still be important effects in \bsbsbar\
mixing. Note also that, due to the mass suppression, the FCNS contribution
to decays involving lighter fermions is completely negligible.

The one decay in which the mass suppression is not obviously a disadvantage
is $B^0_q \to l^+ l^-$, $q=d,s$, since such a suppression is present even
in the SM. We find
\eq
\Gamma(B^0_q \to l^+ l^-)|_{\sss FCNS} = {5 \over 48 \pi} \, 
G_{\sss F}^2 M_{\sss B_q} f_{\sss B_q}^2 m_l^2 \, |F_{qb}|^2 \,
\left( {M_{\sss B_q} \over M_{\sss H^0}} \right)^4 ~.
\eeq
Again taking $M_{\sss H^0} = 100$ GeV, $F_{sb} = 0.17$, and $F_{db} = 0.02$,
this gives
\eq
\eqalign{
BR(B^0_s \to \tau^+ \tau^-)|_{\sss FCNS} & = 2.3 \times 10^{-6} 
~(f_{\sss B_s}/232~{\rm MeV})^2 ~, \cr
BR(B^0_s \to \mu^+ \mu^-)|_{\sss FCNS} & = 8.1 \times 10^{-9} 
~(f_{\sss B_s}/232~{\rm MeV})^2 ~, \cr
BR(B^0_d \to \tau^+ \tau^-)|_{\sss FCNS} & = 2.1 \times 10^{-8} 
~(f_{\sss B_d}/200~{\rm MeV})^2 ~, \cr
BR(B^0_d \to \mu^+ \mu^-)|_{\sss FCNS} & = 7.5 \times 10^{-11} 
~(f_{\sss B_d}/200~{\rm MeV})^2 ~. \cr}
\eeq
For $B_d^0$ decays, the new-physics effects are within the errors of the SM
prediction. For $B_s^0$ decays, the branching ratios due to
flavor-changing neutral scalars are a factor of 2-3 times larger than in
the SM. Since there are uncertainties in the SM prediction, and since we
have taken optimal values for the new-physics parameters, this can only be
considered a marginal signal of new physics.

Therefore, in models with flavor-changing processes mediated by neutral
scalars, there are no ``smoking gun'' signals in penguin decays. For
maximal values of the new-physics parameters, there may be enhancements in
the branching ratios of $b \to s \, \tau^+ \tau^-$ and $B^0_s \to l^+ l^-$.
However, for other values of these parameters there will be no significant
effects in these and other penguin decays, even though there may still be
important contributions to \bqbqbar\ mixing.

\subsection{Supersymmetry}

The parameter space of supersymmetric models is quite complex, so that
definite predictions of effects in penguin decays are hard to obtain. For
example, consider the decay $b \to s \gamma$ in the MSSM \btosgamma. In
addition to the charged-Higgs effects, which always increase the rate [see
Sec.~{\it 2.3(i)}], there are additional SUSY contributions from charginos
$+$ up-type squarks or gluinos $+$ down-type squarks in the loop. In
certain regions of parameter space the net effect is to cancel the
contribution of the $H^\pm$, resulting in a branching ratio at or below the
SM value, while in other regions the branching ratio is always greater than
in the SM. Thus, the experimental measurement of $b\to s\gamma$ does not
constrain the SUSY parameter space in a simple way.

\ref\SUSYbsll{P. Cho, M. Misiak and D. Wyler, \prd{54}{96}{3329}.}

The decay $b\to s \, l^+l^-$ has recently been analysed in supersymmetric
models, taking into account the constraint from $b\to s\gamma$ \SUSYbsll. 
In the MSSM, it is found that the branching ratios for $b\to s \, e^+e^-$
and $b\to s \, \mu^+\mu^-$ can be changed by at most 23\% and 12\%,
respectively. These deviations are within the errors of the SM predictions,
so these decay modes cannot be used to detect supersymmetry. However, the
authors of Ref.~\SUSYbsll\ also study the C-odd lepton-antilepton energy
asymmetry:
\eq
{\cal A} = {N(E_{l^-} > E_{l^+}) - N(E_{l^+} > E_{l^-}) \over
	 N(E_{l^-} > E_{l^+}) + N(E_{l^+} > E_{l^-})}~,
\eeq
in which $N(E_{l^-} > E_{l^+})$ denotes the number of decays where the
$l^-$ is more energetic in the $B$ meson rest frame than the $l^+$. They
find that this asymmetry can be affected by up to 70\% for $b\to s \,
e^+e^-$ and 48\% for $b\to s \, \mu^+\mu^-$. Furthermore, these sizeable
deviations occur in a large region of SUSY parameter space. Thus, this
asymmetry is an excellent place to look for effects of supersymmetry.
Unfortunately, it is not clear how that region of parameter space which
leads to large deviations in this asymmetry is correlated with that region
of parameter space in which there are significant contributions to
\bqbqbar\ mixing.

Ref.~\SUSYbsll\ also examines $b\to s \, l^+l^-$ in a certain class of
nonminimal SUSY models. In this case, the effects can be huge: the
branching ratios for $b\to s \, e^+e^-$ and $b\to s \, \mu^+\mu^-$ can be
doubled, and the asymmetries enhanced by a factor of 3.

For the decays $B_q^0 \to l^+l^-$ and $b\to q q' {\bar q}'$, similar
studies have not yet been carried out in the context of supersymmetric
models. However, since the branching ratio for $b\to s \, l^+l^-$ is not
substantially affected in the MSSM, this suggests that the SUSY
contributions to $B_q^0 \to l^+l^-$ and $b\to q q' {\bar q}'$, which are
similar decays, will also not change the branching ratios significantly. On
the other hand, in nonminimal SUSY models, the branching ratios for these
decays may receive important corrections.

\section{Summary}

The phase information of the CKM matrix, which is the SM explanation of CP
violation, is represented by the unitarity triangle. At present, our
knowledge of this triangle is rather poor -- only the sides have been
measured directly ($|V_{ub}/V_{cb}|$ is probed in charmless $B$ decays, and
within the SM $|V_{td}/V_{cb}|$ can be extracted from \bdbdbar\ mixing),
but these measurements suffer from large theoretical uncertainties. In the
near future, the angles of the unitarity triangle will be extracted from CP
asymmetries in $B$-meson decays. Through the measurements of the CP angles
$\alpha$, $\beta$ and $\gamma$, it will be possible to test the consistency
of this description. There are three distinct ways in which the presence of
new physics might be revealed:

\topic{(1)} The relation $\alpha+\beta+\gamma=\pi$ is violated. (Note that
this relation can be tested only if $\gamma$ is measured in CP asymmetries
involving $B_s^0$ decays. If $\gamma$ is measured via $B^\pm \to\dcp\
K^\pm$ this relation will hold even in the presence of most types of new
physics.) 

\topic{(2)} Although $\alpha+\beta+\gamma=\pi$, one finds values for the CP
phases which are outside of the SM predictions.

\topic{(3)} The CP angles measured are consistent with the SM predictions,
and add up to $180^\circ$, but are inconsistent with the measurements of
the {\it sides} of the unitarity triangle.

\endtopic

If any of these discrepancies is found, we will want to know what kind of
new physics is involved. The principal way in which new physics can affect
the CP asymmetries is through new contributions to \bqbqbar\ mixing. By
performing a model-by-model analysis of physics beyond the SM, it is
possible to ascertain which types of new physics might be responsible for
the discrepancy. This analysis allows us to separate new-physics models
into two types: (i) those in which the phase of \bqbqbar\ mixing is
changed, in which case each of items {\it (1)}-{\it (3)} may occur, and
(ii) those in which the phase is unchanged, in which case only {\it (3)} is
possible. However, this analysis does not tell us how to distinguish among
models of a given type. It is this question which we have attempted to
address in this paper.

Our main observation is quite simple. Any new physics which contributes to
\bdbdbar\ or \bsbsbar\ mixing will necessarily contribute to the rare
flavor-changing penguin decays $b\to dX$ and $b\to sX$. In some cases, the
values of the new-physics parameters which yield significant effects in
\bqbqbar\ mixing will also lead to large deviations from the SM predictions
for certain penguin decays. It is therefore possible to partially
distinguish among different models of new physics by examining their
predictions for these penguin decays. 

\topic{(i) Four generations} The phase of \bqbqbar\ mixing can be changed
due to the new box-diagram contributions with internal $t'$ quarks.
Although the new-physics parameter space is too large to make absolute
predictions for branching ratios of penguin decays, we have shown that
there are regions of parameter space in which both $B$ CP asymmetries and
$B$ penguin decays are significantly affected. The particular example we
chose found important contributions to \bdbdbar\ mixing, and roughly an
order-of-magnitude enhancement of the branching ratios for both exclusive
$b\to d$ electroweak penguins and $B_d^0\to l^+l^-$.

\topic{(ii) $Z$-mediated flavor-changing neutral currents} The phase of
\bqbqbar\ mixing can be altered due to the tree-level exchange of a $Z$
with flavor-changing couplings. In fact, \bdbdbar\ mixing can be dominated
by this new physics. If these new contributions are important, there will
also be unmistakable effects in $b\to q \, l^+ l^-$, $B^0_q \to l^+ l^-$,
and electroweak penguins. In particular, the rates for $b\to s$ ($b\to d$)
penguin processes can be enhanced by as much as one (two) orders of 
magnitude. In addition, $Z$-mediated FCNC's may lead to direct CP violation
in decay modes such as $B^+ \to \pi^+\pi^0$. On the other hand, if the
branching ratios for $b\to q \, l^+ l^-$ are found to be consistent with
the SM, this will indicate that $Z$-mediated FCNC effects in \bqbqbar\
mixing are negligible.

\topic{(iii) Multi-Higgs-doublet models with natural flavor conservation}
There are new box-diagram contributions to \bqbqbar\ mixing with internal
charged Higgs bosons, but the phase of this mixing is unchanged. (If one
also has spontaneous CP violation in such models, this phase is zero, since
the CKM matrix is real, and the unitarity triangle becomes a straight
line.) For that region of parameter space in which \bqbqbar\ mixing is
dominated by the charged-Higgs contribution, the branching ratios for
$B_q^0 \to l^+l^-$  and $B \to X_q \, l^+ l^-$ are enhanced by up to an
order of magnitude or a factor of 2, respectively. However, when the box
diagrams with internal $W^\pm$ and $H^\pm$ bosons are about equal in
magnitude, charged-Higgs effects in penguin decays may not be sufficiently
large to be detected, due to theoretical uncertainties in the SM rate
calculations.

\topic{(iv) Multi-Higgs-doublet models with flavor-changing neutral
scalars} The phase of \bqbqbar\ mixing can be changed due to the tree-level
exchange of a neutral scalar with flavor-changing couplings. However,
there are no significant effects in $B$ penguin decays. This is due to the
fact that the flavor-conserving coupling of the neutral scalar to a
fermion is proportional to the fermion mass, and penguin decays all involve
light fermions.

\topic{(v) Left-right symmetric models} There are no significant effects in
the $B$ system in these models. The only exception is the case where the
right-handed CKM matrix is considerably fine-tuned, but we do not consider
this possibility.

\topic{(vi) Minimal supersymmetric models} There are many new contributions
to \bqbqbar\ mixing, but all have the same phase as in the SM. A search of
the parameter space reveals that SUSY contributions to $b\to s \, l^+l^-$
do not change its branching ratio significantly. This suggests that the
branching ratios for the decays $B_q^0 \to l^+l^-$ and $b\to q q' {\bar
q}'$ will also be relatively unaffected. However, SUSY can be detected by
examining the lepton-antilepton energy asymmetry. This asymmetry can be
affected by up to 70\% for $b\to s \, e^+e^-$ and 48\% for $b\to s \,
\mu^+\mu^-$. 

\topic{(vii) Nonminimal supersymmetric models} In nonminimal SUSY models
with quark-squark alignment, the SUSY contributions to \bqbqbar\ mixing
(and hence to $B$ penguin decays) are generally very small (though models
do exist in which $M_{12}^{\sss SUSY}(B_d^0)/M_{12}^{\sss SM}(B_d^0)
\approx 0.15$). In alternative nonminimal SUSY ``models" it is simply
assumed that all SUSY parameters take the maximum allowed values allowed by
experiment. These models are not terribly predictive, due to the very large
numbers of parameters. It is possible to find new contributions to the
mixing with different phases than in the SM, and to arrange the many
parameters such that the branching ratios of penguin decays are enhanced or
suppressed. However, it is virtually impossible to analyse the effects of
nonminimal SUSY on \bqbqbar\ mixing and $B$ penguins in any systematic
way. 

\endtopic

If some indication of new physics is found in the measurements of CP
asymmetries, the above analysis may be used to distinguish different
candidate models of physics beyond the SM. (In fact, in some cases it is
likely that the new physics will be found first through measurements of
rare $B$ decays.) For example, suppose that new physics is found through a
discrepancy of type {\it (1)} or {\it (2)}.  This would indicate that the
new physics is probably either a fourth generation, $Z$-mediated FCNC's, or
flavor-changing neutral scalars. Since each of these three models affects
$B$ penguin decays differently, they can be at least partially
differentiated by a study of such decays. And if the new physics is found
through a discrepancy of type {\it (3)}, the new physics is likely to be
either a multi-Higgs-doublet model with NFC, or the MSSM. In this case, it
may be difficult to distinguish the two models of new physics since their
effects on $B$ penguin decays are similar. This is not surprising, since
the MSSM contains two Higgs doublets. Still, there are signals, such as the
lepton-antilepton energy asymmetry in $b\to s \, l^+l^-$, which can
differentiate these two models.

To sum up, most physics beyond the SM which can affect CP asymmetries in
$B$ decays will also contribute to rare, flavor-changing $B$ decays. We
have examined the effects of a number of models of new physics on both CP
asymmetries and penguin decays. Although not all models of new physics have
``smoking gun'' signatures in these decays, we have shown that the
measurements of CP asymmetries and rare penguin decays give complementary
information, and both will be necessary if we hope to identify the new
physics.

\bigskip
\centerline{\bf Acknowledgements}
\bigskip

This research was financially supported by NSERC of Canada and FCAR du
Qu\'ebec, by the United States--Israel Binational Science Foundation under
Research Grant Agreement 94-00253/1, and by the Israel Science Foundation.
DL is grateful for the pleasant hospitality of the Physics Department at
the Technion, and of GTAE at the Instituto Superior T\'ecnico in Lisbon,
where some of this work was done. MG would like to thank the Benasque
Center for Physics, at which part of this work was done, for a congenial
atmosphere. We would also like to thank J. Bernabeu, S. Bertolini, G.
Branco, J. Cline, G. Eilam, Y. Grossman, L. Lavoura, E. Nardi and Y. Nir
for helpful conversations. We greatly appreciate the comments of Y.
Grossman, Y. Nir and J. Rosner on the first draft of the manuscript.

\listrefs
\bye